%% file: paper.tex
\documentclass[11pt,a4paper]{article}
\pdfoutput=1   

\usepackage{alpha}
\hypersetup{%
   pdftitle    = {Slow running of the Gradient Flow coupling from 200 MeV to 4 GeV in Nf=3 QCD},
   pdfauthor   = {M. Dalla Brida, P. Fritzsch, T. Korzec, A. Ramos, S. Sint, R. Sommer},
   pdfkeywords = {Lattice QCD, Renormalization group}
}%
\usepackage{macros_alpha}
\usepackage{subcaption}
\usepackage{caption}
\usepackage[section]{placeins}
\usepackage{defs}

\begin{document}

\input{title.tex}

\section{Introduction}
\input{intro.tex}

\section{The running coupling}
\label{sec:coupling}
\input{coupling.tex}

\section{General considerations on the continuum limit of flow
  quantities}
\label{sec:cont}
\input{continuum.tex}

\section{Continuum extrapolations and the $\beta$-function}
\label{sec:step}
\input{step.tex}

\section{Connection of scales $1/L_0$ and $1/L_\mathrm{had}$}
\label{sec:scales}
\input{matchL0.tex}
\label{sec:L0}

\section{Discussion}
\label{sec:conclusions}

\input{conclusions.tex}

\addcontentsline{toc}{section}{Acknowledgement}
\begin{acknowledgement}
We thank our colleagues in the ALPHA collaboration, in particular
C.~Pena, S.~Schaefer, H.~Simma, and U.~Wolff for many useful discussions.

  We would also like to show our gratitude to S. Schaefer and H. Simma
  for their 
  invaluable contribution regarding important modifications to the
  \texttt{openQCD} code. 

Furthermore, we have benefited from the joint production of gauge
field ensembles with a project 
computing the running of quark masses. We thank I. Campos, C. Pena and
D. Preti for this 
collaboration.
We also thank Pol Vilaseca who computed the used one-loop
coefficient of $\cttil$.

We thank the computer centres at HLRN (bep00040) and NIC at DESY, Zeuthen for
providing computing resources and support. 
We are indebted to Isabel Campos and thank her and the staff at the
University of Cantabria at IFCA in the Altamira HPC facility
for computer resources and technical support.

R.S. acknowledges support by the Yukawa Institute for Theoretical
Physics at Kyoto University, where part of this work was carried out.
S.S.  acknowledges support by SFI under grant 11/RFP/PHY3218.  
P.F. acknowledges financial support from the Spanish MINECO's ``Centro de
Excelencia Severo Ochoa'' Programme under grant SEV-2012-0249, as well
as from the grant FPA2015-68541-P (MINECO/FEDER). 
This work is based on previous work \cite{Sommer:2015kza} supported strongly by
the Deutsche Forschungsgemeinschaft in the SFB/TR~09. 
\end{acknowledgement}

\newpage
\appendix

\section{Simulation details}
\label{ap:simulation}

\input{app_tuning.tex}



\bibliography{final}

\end{document}

%% file: title.tex
%
%
\preprintno{%
CERN-TH-2016-160\\
DESY 16-133\\
IFT-UAM/CSIC-16-067\\
WUB/16-02\\
YITP-16-89\\
}
%
%
\title{{Slow running of the Gradient Flow coupling \\ 
from 200 MeV to 4 GeV  in  $\nf=3$ QCD }}
%
%
\collaboration{\includegraphics[width=2.8cm]{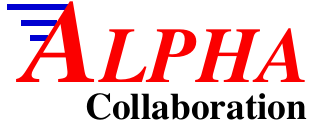}}
%
%
\author[desy]{Mattia~Dalla~Brida}
\author[uam]{Patrick~Fritzsch} %
\author[wup]{Tomasz~Korzec} %
\author[cern]{Alberto~Ramos} 
\author[dublin]{Stefan~Sint}%
\author[desy,hu]{Rainer~Sommer}
%
%
\address[desy]{John von Neumann Institute for Computing (NIC), DESY, Platanenallee~6, 15738 Zeuthen, Germany}
\address[uam]{Instituto de F\'{\i}sica Te{\'o}rica UAM/CSIC, Universidad Aut{\'o}noma de Madrid,\\
C/ Nicol{\'a}s Cabrera 13-15, Cantoblanco, Madrid 28049, Spain}
\address[wup]{Department of Physics, Bergische Universit\"at Wuppertal, Gau\ss str. 20,
42119 Wuppertal, Germany}
\address[cern]{CERN, Theory Division, Geneva, Switzerland}
\address[dublin]{School of Mathematics, Trinity College Dublin, Dublin
  2, Ireland}
\address[hu]{Institut~f\"ur~Physik, Humboldt-Universit\"at~zu~Berlin, Newtonstr.~15, 12489~Berlin, Germany}

%

%
\begin{abstract}
\noindent\rule{\textwidth}{0.4pt}\\[0.6\baselineskip] 
Using a finite volume Gradient Flow (GF) renormalization scheme with
Schr\"odinger Functional (SF) boundary conditions, we compute the
non-perturbative running coupling in the range $2.2 \lesssim
{\bar g}_\mathrm{GF}^2(L) \lesssim 13$.  Careful continuum extrapolations turn out to
be crucial to reach our high accuracy. The running of the coupling is always
between one-loop and two-loop and very close to one-loop in the region of
$200\,\MeV \lesssim \mu=1/L \lesssim 4\,\GeV$. While there is no convincing
contact to two-loop running, we match non-perturbatively to the
SF coupling with background field. In this case we know the
$\mu$ dependence up to $\sim 100\,\GeV$ and can thus connect to the
$\Lambda$-parameter.\\
\rule{\textwidth}{0.4pt}\\[\baselineskip] 
\end{abstract}
%
%
\begin{keyword}
Lattice QCD \sep Renormalization group 
\PACS{%
12.38.Gc   
}
\end{keyword}

\maketitle

\tableofcontents

\newpage


%% file: intro.tex
The energy dependence of the strong coupling constant $\alpha_s(\mu)$
in a physical scheme provides information on how to connect the low
and high energy regimes of 
QCD. Relating these very different domains of the strong interactions
is key to providing a solid determination of the fundamental parameters
of the Standard Model~\cite{Brida:2016flw}. Lattice QCD is in
principle an ideal 
tool for such studies. Observables defined at short Euclidean distances
can be used for a non-perturbative physical coupling definition (see for
example~\cite{Sommer:1997xw} and references cited therein), and
its value can be extracted accurately via Monte Carlo simulations. A
direct implementation of this program has to face
the so-called window problem: the short Euclidean distance used to
define the renormalization scale has to be both large compared to
the lattice spacing $a$ and small compared to the
total size of the box (denoted by $L$) used in the
simulation. Since the box has to be large enough to describe
hadronic physics, computational constraints severely limit
the range of renormalization scales that one can study. 

\emph{Finite size scaling} provides an elegant solution for this 
problem~\cite{Luscher:1991wu}. Relating the
renormalization scale $\mu$ with the finite size of the box via $\mu =
1/L$, the coupling $\gbar^2(L)$ depends on only one
scale.\footnote{We use a massless renormalization
    scheme. Renormalized couplings and renormalization factors of
    quark masses and composite operators are defined at zero quark
    mass, and the renormalization
group functions do not depend on the quark masses.} 
Lattices of different volumes can be matched, allowing us to compute the 
\emph{step scaling function} $\sigma(u)$~\cite{Luscher:1991wu}. It 
measures how much the coupling changes when the renormalization scale
changes by a fixed factor, which we set to two,
\begin{equation}
  \sigma(u) = \gbar^2(2L) \Big|_{\gbar^2(L) = u}\,.
  \label{e:sigmaformal}
\end{equation}
It can be considered a discrete version of the renormalization
group $\beta$-function. The exact relation is 
  \begin{equation}
    \log(2) = 
    -\int_{\sqrt{u}}^{\sqrt{\sigma(u)}} \frac{\rmd x}{\beta(x)} \,,
  \end{equation}
with the convention
\begin{equation}
  \beta(\gbar) = - L \frac{\partial \gbar(L)}{\partial L} 
\underset{\gbar\rightarrow 0}{\sim} -b_0 \gbar^3 - b_1 \gbar^5 + \ldots\,,
\end{equation}
where the universal coefficients in the asymptotic expansion take the values
$b_0 = 9/(16\pi^2)$ and $b_1 = 1/(4\pi^4)$ in $\Nf=3$ QCD. 

Once $\sigma(u)$ is known one can set $u_0 =
\bar g^2(L_0)$ and use the 
recursive relation
\begin{equation}
  u_k = \sigma(u_{k-1}),\qquad k=1,\dots,n_s, \label{e:recurs}
\end{equation}
to relate non-perturbatively the scale $1/L_0$ with the 
scales $2^{-k}/L_0$ for $k=0,\dots,n_s$. A few iterations suffice
to connect a hadronic low energy scale with the electroweak scale. 

This is the strategy of the ALPHA collaboration.
Using the so called
Schr\"odinger Functional (SF) scheme~\cite{Luscher:1992an,Sint:1993un},
QCD with $\nf=0,2$ and $\nf=4$ quark flavors has been
studied~\cite{Luscher:1993gh,DellaMorte:2004bc,Tekin:2010mm}. 
Of immediate relevance to the present work is the 
recent application of this technique to the high 
energy domain of $\nf=3$ QCD~\cite{Brida:2016flw}. There
the energy dependence of the strong coupling was studied between the 
electroweak scale and an intermediate energy 
scale $\mu_0 = 1/L_0 \sim 4\,{\rm GeV}$, defined by $\bar g^2_{\rm SF}(L_0)=2.012$,
with very high accuracy.

This strategy is theoretically very appealing, but has some practical
difficulties. The computational   
cost of measuring the SF coupling grows fast at low
energies and in particular towards the continuum limit. Thus it is 
challenging to reach the low energy domain
characteristic of  
hadronic physics, especially if one aims at maintaining the high
precision achieved in~\cite{Brida:2016flw}. The recently proposed
coupling definitions 
based on the Gradient Flow (GF)~\cite{Luscher:2010iy} are much better
suited for this task. The relative precision of the GF
coupling in a Monte Carlo
simulation is typically high and shows a weak dependence on
both the energy scale and the cutoff (see~\cite{Ramos:2015dla} for a recent
review and more quantitative statements). Moreover GF couplings
can easily be
used in combination with \emph{finite size scaling} and a particular
choice of boundary
conditions~\cite{Fodor:2012td,Fritzsch:2013je,Ramos:2014kla,Luscher:2014kea}.   

In this work we use the GF coupling defined with 
SF boundary
conditions~\cite{Fritzsch:2013je} (denoted by $\bar g^2_{\rm
  GF}(L)$) to connect non-perturbatively the intermediate energy
scale $1/L_0$ with a typical hadronic scale $1/L_{\rm had}$ defined by
the condition
\begin{equation}
  \bar g^2_{\rm GF}(L_{\rm had}) = 11.31\,.
\end{equation}
The main result of this paper is the relation 
\begin{equation}
  L_{\rm had} = 21.86(42)\, L_0\,. \label{e:scalefactor}
\end{equation}
As the reader will see, our choices of lattice discretization and scale
$1/L_{\rm had}$ are such that $L_{\rm had}$ can be related
with the pion and kaon decay constants by using the \texttt{CLS}
ensembles~\cite{Bruno:2014jqa}. This work therefore represents an
essential step in the ALPHA collaboration effort of a first principles
determination of the strong coupling constant and quark
masses at the electroweak scale in terms of low energy hadronic
observables~\cite{Brida:2015gqj,Campos:2015fka}.  

The paper is organized as follows. In section~\ref{sec:coupling} we
fix our notation and introduce the details of our coupling
definition. \Sect{sec:cont} discusses general aspects of taking the
continuum limit while \sect{sec:step} contains the extraction 
of the continuum $\sigma(u)$. After arriving at our main result
in \sect{sec:scales} we  discuss our findings in \sect{sec:conclusions}.


%% file: coupling.tex
\subsection{Continuum}

We work in 4-dimensional Euclidean space and 
consider standard SF boundary conditions with
zero background 
field~\cite{Luscher:1992an,Sint:1993un}. In summary, gauge fields
are periodic in the three spatial directions with period $L$, and the
spatial components $k=1,2,3$ of the gauge field satisfy homogeneous 
Dirichlet boundary conditions in time, 
\begin{equation}
  A_k(0,\vecx)=A_k(T,\vecx) = 0\,. 
\end{equation}
Fermion fields are required to obey periodic boundary
  conditions in space up to a phase,
  \begin{equation}
    \begin{split}
      \psi(x+L\hat k) &= e^{\imath \theta}\psi(x)\,,\\
      \psibar(x+L\hat k) &= e^{-\imath \theta}\psibar(x)\,. \\ 
    \end{split}
  \end{equation}
We choose the value $\theta =1/2$~\cite{Sint:1998iq}. 
Defining the projectors $P_{\pm} = \frac{1}{2}(1\pm \gamma_0)$,
the time boundary conditions read
\begin{equation}
\label{eq:fermionbc}
  \begin{split}
    P_+\psi(0,\vecx) = &\,0 =\, \bar \psi(0,\vecx)P_- \,,\\
    P_-\psi(T,\vecx) = &\, 0 =\, \bar \psi(T,\vecx)P_+ \,.\\
  \end{split}
\end{equation}
The GF~\cite{Narayanan:2006rf,Luscher:2010iy} defines a
family of gauge fields $B_\mu(t,x)$ parametrized by the flow time $t\geq0$
via the equation\footnote{Unless stated otherwise, repeated Greek
  indices are summed from 0 to 3. Repeated Latin indices are either
  summed from 1 to 8 ($a,b,\ldots$) or from 1 to 3 ($i,j,\dots$).}
\begin{equation}
  \label{eq:flow}
  \partial_t B_\mu(t,x) = 
  D_\nu G_{\nu\mu}(t,x)\,, \qquad B_\mu(0,x) = A_\mu(x)\,,
\end{equation}
where $D_\mu = \partial_\mu + [B_\mu, \cdot]$ is the covariant
derivative, and $G_{\mu\nu}(t,x)$ is the field strength tensor of
the flow field,
\begin{equation}
  G_{\mu\nu} = \partial_\mu B_\nu - \partial_\nu B_\mu + [B_\mu,B_\nu]\,.
\end{equation}
Gauge invariant composite operators defined from the flow field
$B_\mu(t,x)$ are renormalized
observables, see~\cite{Luscher:2011bx}. In particular, our 
definition of a running coupling follows the proposal 
 of using the action density at positive flow
time~\cite{Luscher:2010iy}. In a finite volume and with our choice of boundary conditions
the running coupling was defined
in~\cite{Fritzsch:2013je}  
\begin{equation}
  \label{eq:coupling}
  \bar g^2_{\rm GF}(L) = \mathcal N^{-1}(c)\frac{t^2}{4}
  \frac{\langle G_{ij}^a(t,x)G_{ij}^a(t,x)\,
    \delta_{Q,0}\rangle}{\langle \delta_{Q,0} \rangle}
  \Big|_{\sqrt{8t}=cL,\,x_0=T/2} \,,
\end{equation}
where $\mathcal N(c)$ is a known function~\cite{Fritzsch:2013je}. Note
that we use only the spatial components of the field strength
tensor to define the coupling. As argued
in~\cite{Fritzsch:2013je} 
boundary effects are smaller for this particular coupling definition,
while we have observed that one does not lose numerical
precision. The coupling is defined by projecting to
the sector of vanishing topological charge, 
$Q=\frac{1}{32 \pi^2} \int_x \epsilon_{\mu\nu\rho\sigma}
G_{\mu\nu}^a(t,x) G_{\rho\sigma}^a(t,x)$, via the insertion of
$\delta_{Q,0}$ into the path integral expectation values. 
This choice is convenient because lattice
simulations with SF boundary conditions suffer from the topology
freezing problem at small lattice
spacing~\cite{DelDebbio:2004xh,Schaefer:2010hu,Fritzsch:2013yxa}. Projecting to the zero charge
sector avoids this problem~\cite{Fritzsch:2013yxa}.  
The renormalization scheme is completely defined by adding
that we use 
\begin{equation}
T=L \;\text{ and }\; c=0.3\,.
\end{equation}
This choice is fixed in this work, apart from
\sect{sec:cont} where we also consider
other values of $c$.

\subsection{Lattice}

For our lattice computations we work on a $(L/a)^3\times (T/a)$
lattice with lattice spacing $a$. We use the tree-level improved
Symanzik gauge 
action~\cite{Luscher:1985zq}. With $\mathcal S_0$ and $\mathcal S_1$
denoting the set of $1\times 1$ and $2\times 1$ oriented loops
respectively, we have
\begin{equation}
  \label{eq:Sgauge}
  S_\mathrm{G}[U] = \frac{1}{g_0^2}\sum_{k=0}^1 c_k \sum_{\mathcal C \in \mathcal
  S_k} w_k(\mathcal C){\rm tr} [1 - U(\mathcal C)]\,,
\end{equation}
where $U(\mathcal C)$ denotes the product of the link variables
$U_\mu(x)$ around the loop $\mathcal C$. Tree-level $\rmO(a^2)$
bulk improvement is guaranteed 
by choosing $c_0=5/3$ and $c_1=-1/12$. 
Modifications of the gauge action near the
time boundaries $x_0=0,T$ lead to Schr\"odinger Functional boundary
conditions in the continuum.

We stick to {\sf option B} of
reference~\cite{Aoki:1998qd} and choose the weights $w_k(\mathcal C)$
as follows:\footnote{All simulations were performed with a modified
  version of the \texttt{openQCD v1.0}
  package~\cite{Luscher:2012av}. The documentation of the
  package provides useful information for the interested reader.}
\begin{subequations}
\begin{equation}
  w_0(\mathcal C) = \left\{
    \begin{array}{ll}
      1/2, & \text{all links in $\mathcal C$ are on the time boundary}\\
      \ct(g_0), & \text{$\mathcal C$ has one link on the time boundary}\\
      1, & \text{otherwise}\\
    \end{array}
  \right.\,,
\end{equation}
\begin{equation}
  w_1(\mathcal C) = \left\{
    \begin{array}{ll}
      1/2, & \text{all links in $\mathcal C$ are on the time boundary}\\
      3/2, & \text{$\mathcal C$ has two links on the time boundary}\\
      1, & \text{otherwise}\\
    \end{array}
  \right.\,.
\end{equation}
\end{subequations}
The improvement coefficient $\ct$ is inserted with the available one-loop 
precision, see \sect{s:boundary}.
We simulate three massless flavors of non-perturbatively
$\rmO(a)$-improved Wilson fermions 
with action
\begin{equation}
  \label{eq:Sfermion}
  S_\mathrm{F}[U,\overline\psi,\psi] = a^4 \sum_{i=1}^{\nf} \sum_{x}
  \overline \psi_i(x) (D+m_0)\psi_i(x)  \,, 
\end{equation}
where $m_0$ is the bare quark mass that we set to the critical
  value $\mcrit$. The Dirac operator can
be decomposed as
\begin{equation}
  D = D_\mathrm{w} + \delta D_{\rm sw} + \delta D_{\rm bnd}, \\
\end{equation}
where $D_\mathrm{w}$ is the usual lattice Wilson--Dirac operator, 
\begin{equation}
  \delta D_{\rm sw} \psi(x) =  ac_{\rm sw}\frac{\imath}{4} 
  \sigma_{\mu\nu}F_{\mu\nu}^{\rm cl}(x)\,\psi(x),
\end{equation}
is the Sheikholeslami--Wohlert term~\cite{Sheikholeslami:1985ij} 
with $F_{\mu\nu}^{\rm cl}$ being the 
lattice clover discretized version of the field strength tensor, and finally
\begin{equation}
  \delta D_{\rm bnd}\psi(x) = (\cttil-1)\frac{1}{a} (\delta_{x_0/a,1} +
  \delta_{x_0/a,\;T/a-1})  \psi(x),
\end{equation}
is the contribution of the fermionic boundary
counterterm~\cite{Luscher:1996sc}. We use the non-perturbatively
determined $c_{\rm sw}(g_0)$~\cite{Bulava:2013cta}. Except at the time 
boundaries, our action is the same as the one used by the \texttt{CLS}
collaboration~\cite{Bruno:2014jqa}.

With our choice of boundary conditions in time, the complete removal of
$\rmO(a)$ effects requires the knowledge of the boundary improvement 
coefficients $\ct, \cttil$. We use their values
determined in perturbation theory. As an estimate of the uncertainty 
of perturbation theory we use the last known term
in the perturbative series, the one-loop term (cf. \sect{s:boundary}). 
Details will be discussed later. 

SF boundary conditions on the lattice are imposed in complete
analogy to the continuum counterparts. The gauge links obey  
\begin{equation}
  \label{eq:linksbc}
  \left.U_k(x)\right|_{x_0=0,T} = 1\,,\qquad k=1,2,3\,,
\end{equation}
while the fermion boundary conditions remain the
same as in the continuum, eq.~\eqref{eq:fermionbc}.

There is much freedom when translating the GF equation
eq.~\eqref{eq:flow}, and the energy density used to define the
coupling (see eq.~\eqref{eq:coupling}) to
the lattice. Different choices differ only by cutoff effects, but
these can be substantial. A popular choice is the Wilson flow
(no summation over $\mu$)
\begin{equation}
  \label{eq:wflow}
  a^2\left(\partial_t V_\mu(t,x)\right) V_\mu(t,x)^\dagger  = 
  -g_0^2 \partial_{x,\mu} S_{\rm W}[V],\qquad V_\mu(0,x) = U_\mu(x) \,,  
\end{equation}
where $V_\mu(t,x)$ are the links at positive flow time and
$\partial_{x,\mu} S_{\rm W}[V]$ is the force deriving from the
Wilson plaquette gauge action
(i.e. eq.~\eqref{eq:Sgauge} with the choices $c_0=1, c_1=0$). It has
been shown~\cite{Ramos:2015baa} that this choice introduces $\rmO(a^2)$ cutoff effects when integrating the flow equation. They can be
avoided by using the
Symanzik $\rmO(a^2)$ improved ``Zeuthen flow''
equation (no summation over $\mu$)
\begin{equation}
  a^2\left(\partial_t V_\mu(t,x)\right) V_\mu(t,x)^\dagger  = 
  -g_0^2 \left(1 + 
    \frac{a^2}{12}\Delta_\mu  
  \right) \partial_{x,\mu} S_{\rm LW}[V],\qquad V_\mu(0,x) = U_\mu(x) \,,  
\end{equation}
where 
$\partial_{x,\mu} S_{\rm LW}[V]$ is the force deriving from the
Symanzik tree-level improved (L\"uscher-Weisz) gauge action
eq.~\eqref{eq:Sgauge} 
(see~\cite{Ramos:2015baa} for more details). 
We insert the correction term
$\Delta_\mu=\nabla_\mu^\ast\nabla_\mu^{}$ into the flow equation for
all links 
$(x,\mu)$ except for those links $(x,0)$ where an end-point 
touches one of the SF boundaries $x_0=0,T$. For those links we 
simply choose~$\Delta_0=0$.

The discretized observable is defined to be 
the action density derived from the L\"uscher-Weisz 
gauge action (i.e. eq.~\eqref{eq:Sgauge}). Our choices guarantee
that, neglecting small terms coming from the time boundaries at
$x_0=0,T$, we do not introduce any $\rmO(a^2)$ cutoff effects neither
through the flow equation nor through the definition of the
observable. The remaining
cutoff effects in our flow quantities are hence produced by our lattice
action eqs.~(\ref{eq:Sgauge},\ref{eq:Sfermion}) and by the initial
condition for the flow equation at $t=0$~\cite{Ramos:2015baa}. Although
this is our 
preferred setup, in several parts 
of the work we will compare the results with the more standard Wilson
flow / clover-observable discretization. 

At non-zero $a/L$ our coupling definition reads
\begin{equation}
  \label{eq:coupling1}
  \gbar^2_{\rm GF}(L) = \gbar^2_{0.3}(L)
\end{equation}
with  
\begin{equation}
 \gbar^2_{c}(L) = t^2\hat{\mathcal N}^{-1}(c,a/L)
  \frac{\langle E_\mathrm{mag}(t,x)  
    \hat\delta(Q)\rangle}{\langle \,\hat\delta(Q) \rangle}
  \Big|_{\sqrt{8t}=cL,\, x_0=T/2} \,,
  \label{eq:lattcoupling}
\end{equation}
and
\begin{equation}
   E_\mathrm{mag}(t,x) = \frac{1}{4}[G_{ij}^a(t,x)G_{ij}^a(t,x)]^{\rm LW}.
\end{equation}
Several comments are in order. We have chosen to define the 
coupling through just the magnetic part $E_\mathrm{mag}$ of
$E$ since this choice has a lower sensitivity to the
boundary improvement coefficient $\ct$, and because
its (tree-level) $\rmO(a^2)$ improvement does not need any further 
terms\footnote{In contrast, the electric components would
  require additional terms to cancel total derivative contributions
  that do not vanish because of our Schr\"odinger Functional boundary
  conditions~\cite{Ramos:2015baa}. 
}. 
As in \cite{Fritzsch:2013je} the normalization
factor $\hat{\mathcal N}(c,a/L)$ is computed on the lattice with our
choices of discretization (action, flow and observable), such that 
in the relation $\gbar^2_{\rm GF}=g_0^2+\rmO(g_0^4)$ the leading term
has all lattice artifacts removed. 
Due to the fact that we use a tree-level
improved action, and neither the Zeuthen flow equation nor the
L\"uscher-Weisz observable discretization introduce any $\rmO(a^2)$ artifacts, we furthermore have 
\begin{equation}
   \delta(c,a/L) \equiv \frac{\hat{\mathcal N}(c,a/L)}{\mathcal N(c)}
   - 1  = {\rm O}((a/L)^4)\,\,,   
   \label{e:delta}
\end{equation}
i.e. the lattice normalization in fact only corrects 
sub-leading 
$\rmO((a/L)^4)$ terms. Finally, on the lattice one has
to clarify 
what is meant by projecting to zero topology. We
define the topological charge by \cite{Luscher:2010iy}
\begin{equation}
  \label{eq:qtop}
  Q = \frac{1}{32\pi^2}\sum_x  \epsilon_{\mu\nu\rho\sigma}\,
   [G_{\mu\nu}^a(t,x)
  G_{\rho\sigma}^a(t,x)]^{\rm cl}\,, 
\end{equation}
using the clover discretization of the field
strength tensor of $V$. We then set $\sqrt{8t}=cL$ with
$c=0.3$ and use the Zeuthen flow. With this definition the
topological charge   
is not  integer valued but approaches integers close
to the continuum limit.  Therefore,
the Kronecker $\delta_{Q,0}$ of the continuum definition is 
replaced by 
\begin{equation}
  \label{eq:deltaq}
    \hat\delta(Q) = 
    \begin{cases} 
        1\,, & \text{if } |Q|<0.5\\ 
        0\,, & \text{otherwise}\,. 
    \end{cases}
\end{equation}


%% file: continuum.tex
All studies of finite size scaling with the GF
scheme show significant cutoff effects in the extrapolations of the
step scaling function (see~\cite{Ramos:2015dla} and
references therein). In fact one may be concerned not only by 
the leading\footnote{
The $\rmO(a)$ effects from the SF time boundaries will be ignored
  in the following discussion but considered later.} 
$\rmO(a^2)$ effects, but also by the sub-leading higher order
corrections (in the present case, starting at
$\rmO(a^3)$) that might lead to the wrong continuum limit. 

Local composite fields constructed from the flow field have
a natural length scale given by the smoothing 
radius $\sqrt{8t}$,  which
is smaller than $L$ by a factor $c$. Hence the natural expansion parameter
for the cutoff effects is  $\epsilon = a/\sqrt{8t} = a/(cL)$.
A first example is provided by checking the effects at
tree-level. This just amounts to studying $\delta(c,a/L)$, \eq{e:delta}.
In order to get a  more general picture, we consider besides 
our discretization of the flow observable (``Zeuthen flow''),
also the one used in many studies: Wilson flow and  
clover discretization of the energy density~\cite{Luscher:2010iy} 
(for short ``Wilson flow'').
We find that\footnote{Incidentally, inserting both the continuum
  result ${\cal N}(0.3) = 8.74061\times 10^{-3}$ and these
  parameterizations for $c=0.3$ into eq.~(2.20) yields the 
lattice norms $\hat{\cal N}(0.3,a/L)$ for both lattice discretizations
of the GF coupling and all lattice sizes considered (ca.~3-4
significant digits).}  
\begin{equation}
\delta(c,a/L) \sim  
\begin{cases}
 -0.9118 \epsilon^2 + 0.4867 \epsilon^4  & \text{ Wilson flow}
 \\
 \phantom{-0.9118 \epsilon^2 + }   1.7165 \epsilon^4 & \text{ Zeuthen flow} 
\end{cases}
  \,, \quad \epsilon = \frac{a}{cL}= \frac{a}{\sqrt{8t}} \,,
\end{equation}
where $\sim$ holds with corrections of less than $10^{-4}$ for
$\epsilon < 0.33$ and for  
$c$ in a range 0.1--0.4. One may also consider the GF coupling for 
twisted periodic boundary conditions \cite{Ramos:2014kla}; 
the above numbers hardly change. This example not only shows that in
fact the cutoff effects are predominantly a function of
$\epsilon=a/\sqrt{8t}$, but also that 
the contribution of orders higher than $\epsilon^2$ are only at the level of a few percent for $\epsilon < 0.3$. 
There is a clear hierarchy of the different orders at small 
$\epsilon$, say $\epsilon<0.3$. 

Of course one has to study the situation beyond 
tree-level perturbation theory, and in particular the scaling
properties of the lattice approximation to the step scaling function
$\sigma(u)$ of \eq{e:sigmaformal}.
In order to do so, it is useful to consider the general ratio
\begin{equation}
  \label{eq:defratio}
  R_{c,c'}(u,a/L,s)=\left.\frac{\gbar^2_{c}(L)}{\gbar^2_{c'}(sL)}\right|_{\gbar^2_{c}(L)=u}
  \,, 
\end{equation}
that has a natural expansion 
\begin{equation}
  R_{c,c'}(u,a/L,s) = R_{c,c'}(u,0,s) \Big\{ 
    1 + A_{c,c'}(u)[\epsilon^2 - \epsilon'^2] 
   + \ldots \Big\},
\end{equation}
where $\epsilon=a/(cL)$ and $\epsilon'=a/(c'sL)$. The connection
to the standard step scaling function is $\sigma(u) = u\, R^{-1}_{c,c}(u,0,2)$.
It is again worthwhile to first consider tree-level. To this end,
we temporarily replace the normalization $\hat{\mathcal N}(c,a/L)$
by the continuum one, $\mathcal N$, in eq.~\eqref{eq:lattcoupling};
otherwise all cutoff effects  
are removed. With this replacement, the tree-level ratio
$R_{c,c'}(u,0,s)$ is  
to a very good approximation
just a function of $u$ and the product $sc'$,  
while the function $A_{c,c'}(u)$ depends little on $c,c'$.
An inspection of our numerical data shows that this is 
true also at non-vanishing coupling. These properties 
allow us to get insight into the 
scaling properties of the step scaling function 
by considering the case $s=1$ where we can use our full dataset. As we shall
see in section~\ref{sec:step}, we have 5 lattice resolutions 
$L/a=8,12,16,24,32$ at our disposal. Continuum 
extrapolations can involve a change of the
lattice spacing of up to a factor
four. Moreover these ratios can be 
computed even more precisely than the step scaling function, 
since they are evaluated on the same
ensembles and one profits from the statistical correlation of the
data. 

\begin{figure}[t!]
  \centering
  \includegraphics[width=\textwidth]{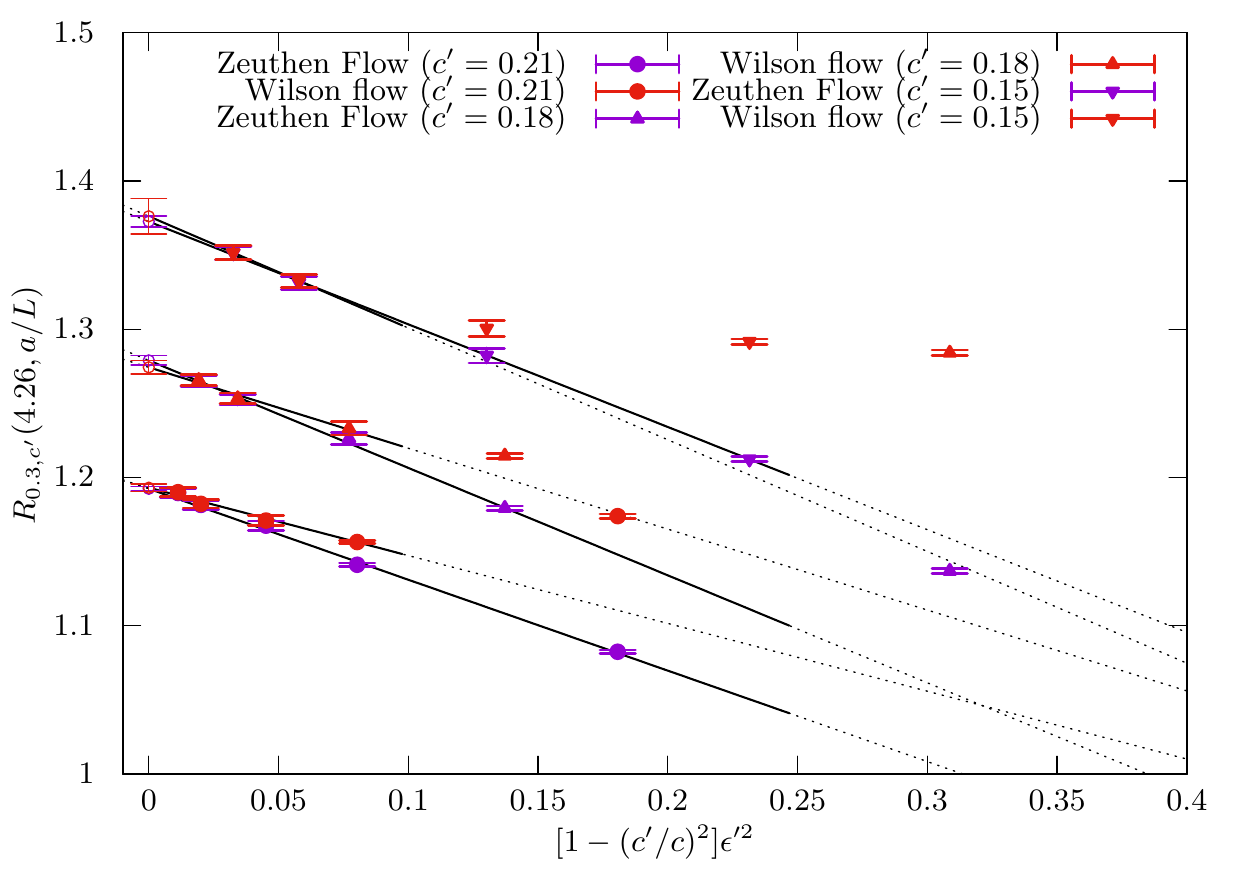}
  \caption{Ratio $R_{0.3,c'}(4.26,a/L)$ for various $c'$ 
  and two different discretizations of the observable. In the definition
  \eq{eq:defratio} all quantities refer to the same discretization. 
  Full lines are linear fits in $a^2$ to data satisfying \eq{e:acuts}.}
  \label{fig:ratio1}
\end{figure}
\begin{figure}[t!]
  \centering
  \includegraphics[width=\textwidth]{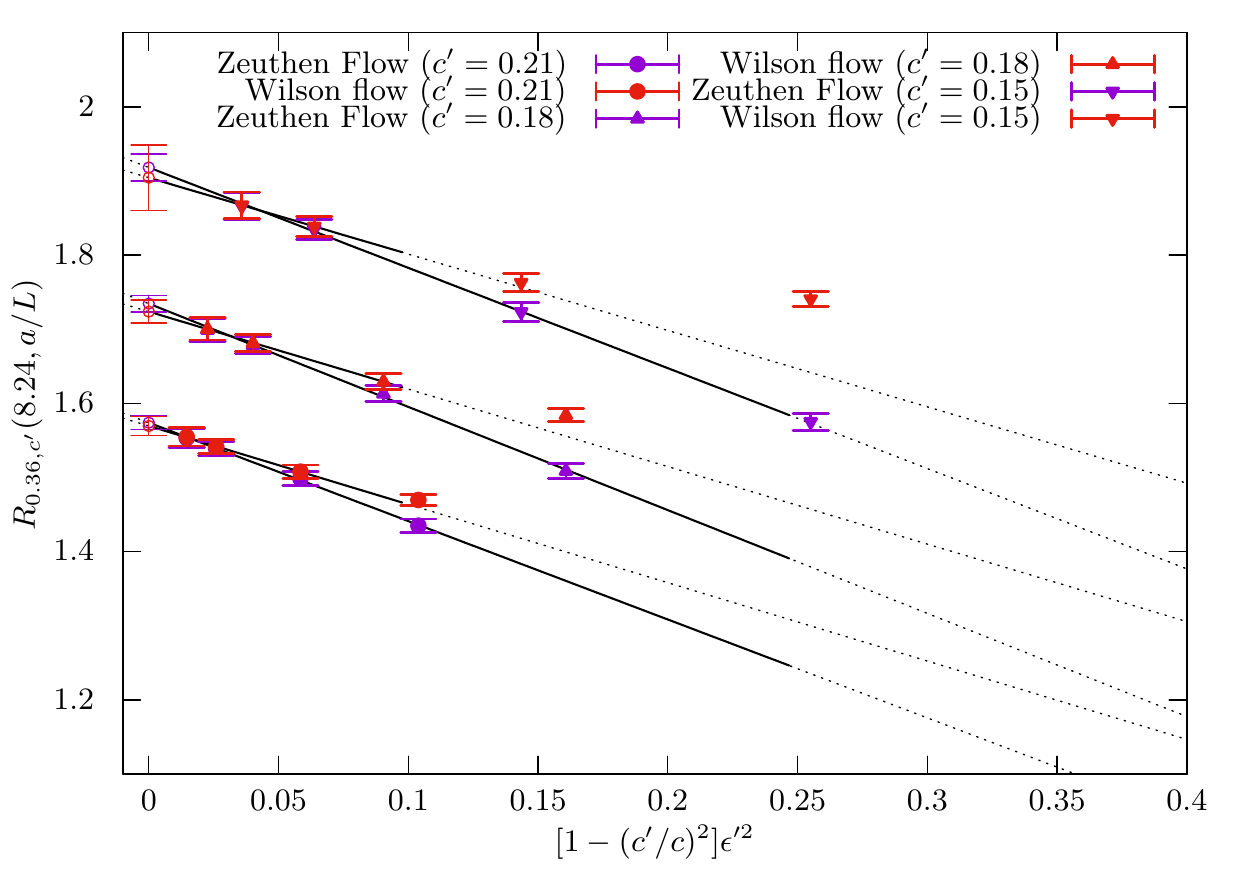}
  \caption{Ratio $R_{0.36,c'}(8.24,a/L)$ for various $c'$ 
  and two different discretizations of the observable. In the definition
  \eq{eq:defratio} all quantities refer to the same discretization. 
  Full lines are linear fits in $a^2$ to data satisfying \eq{e:acuts}.}
  \label{fig:ratio2}
\end{figure}

Figures~\ref{fig:ratio1} and~\ref{fig:ratio2} show $R_{c,c'}(u,a/L,1)$
for  all together six different combinations, $c,c'$ and
two values of $u$. 
The data originate from the simulations described in
\App{ap:simulation}, forming 
first the ratios at the available $\gbar^2_{c}$ and then 
performing a (very smooth) interpolation to the two chosen values of $u$. 
As shown in the figures, 
we separately extrapolate the ratios for the two different discretizations
of the flow observables to the continuum limit. We use a  
pure $a^2$ ansatz for the cutoff effects in the ranges
\begin{equation}
  \begin{split}
    [1-(c'/c)^2]\left( \frac{a}{c'L} \right)^2 &< 0.10 \qquad \text{Wilson flow}\,, \\
    [1-(c'/c)^2]\left( \frac{a}{c'L} \right)^2 &< 0.25 \qquad \text{Zeuthen flow}\,. \\
  \end{split}
  \label{e:acuts}
\end{equation}
The data are compatible with the linear behavior in
$a^2$ and the so-estimated continuum limits agree.
The test is rather stringent because here
the precision is higher than in the step scaling functions,
which form the core observables of the rest of this paper.
For the step scaling functions
there is no analogy of the correlations of numerator and denominator in 
\eq{eq:defratio}, which enhance the precision of
$R_{c,c'}(u,a/L,1)$. \Fig{fig:ratio1} and \fig{fig:ratio2} 
are a good confirmation 
that higher order cutoff effects are small, when \eq{e:acuts}
is satisfied. 

Translating the bounds~\eqref{e:acuts} to the case  
of the step scaling function we have
\begin{equation}
  \begin{split}
    [1-(1/2)^2]\left( \frac{a}{cL} \right)^2 &< 0.10 \qquad \text{Wilson flow}\,, \\
    [1-(1/2)^2]\left( \frac{a}{cL} \right)^2 &< 0.25 \qquad \text{Zeuthen flow}\,. \\
  \end{split}
  \label{eq:acutsssf}
\end{equation}
We then expect the step scaling function computed using
the Zeuthen flow to have only small corrections to an $a^2$ scaling for
$\epsilon^2 = a^2/(cL)^2 < 0.33$. Our coarsest data set
has $L/a=8$ and $c=0.3$, which implies $\epsilon^2 = 0.17$.

The difference in the bounds eq.~\eqref{eq:acutsssf} means
that the more precise continuum limit is obtained for the Zeuthen flow.
Despite the fact that cutoff effects for the Wilson flow are
smaller, their complicated functional form makes extrapolations 
more difficult and less precise. In particular the coarser lattices used
to determine the continuum step scaling function in the next section
would have significant violations of the leading $a^2$ scaling
if we were using the Wilson flow data.

However, one has to state that  the $a^2$ corrections are sizable. 
Since neither the Zeuthen flow equation nor the evaluation of a
classically improved observable introduce any $a^2$ cutoff effects,
these remaining lattice artifacts are a consequence of the  quantum
corrections due to the initial condition of the flow equation at
$t=0$ and due to the action of the fluctuating fields in the 
path integral~\cite{Ramos:2015baa}. Whether there are practical ways
to reduce these remaining $a^2$ effects substantially is an
interesting problem that deserves further attention in the future.  

\subsection{Boundary $\rmO(a/L)$ effects} \label{s:boundary}

With our choice of SF
boundary conditions eqs.~(\ref{eq:linksbc}, \ref{eq:fermionbc}), the
complete removal of $\rmO(a)$ cutoff effects requires not only the
non-perturbative value of the 
coefficient $\csw$~\cite{Bulava:2013cta}, but also the determination of the
boundary coefficients $\ct,\cttil$. These are known only to 
one-loop for our choice of lattice action~\cite{Takeda:2003he,pol,Nf3tuning}
\begin{equation}
  \begin{split}
    \ct &= 1 + \ct^{(1)}g_0^2 + {\rm O}(g_0^4)\,, \qquad
    \ct^{(1)} = 0.0326718\,,
    \\
    \cttil &= 1 + \cttil^{(1)}g_0^2 + {\rm O}(g_0^4)\,,\qquad
    \cttil^{(1)} = -0.01505\,,
    \\
  \end{split}
\end{equation}
and therefore we have to
estimate the possible effects of higher order terms in the coupling.

For this purpose it is
convenient to recall that our GF coupling is 
defined at time-slice $x_0=T/2$, and with our choice $c=0.3$ and $T=L$
the smearing radius is $\sqrt{8t} = cL = 0.3T$. Therefore we expect
boundary effects to be suppressed, since our observable is localized
at the center of the lattice, away from the boundaries. The issue
was investigated in \cite{Luscher:2014kea} with the 
conclusion that indeed these boundary contributions are small. 
Here we estimate the effect quantitatively and specifically for
our observable. 

We first quote the linear $a$-effects at leading order in 
perturbation theory. They are obtained by expanding the 
tree-level norm $\cal N$ in $\ct-1$, treating $\ct = 1 + \rmO(g_0^2)$.
The result is 
\begin{eqnarray}
    \label{e:deltact}
    \Sigma(u,a/L)  &=& \Sigma(u,a/L)_{\ct=1} +  \frac{\ct-1}{g_0^2 \ct^{(1)}} \Delta^{\ct} \Sigma(u,a/L)\,,
    \\
  \Delta^{\ct} \Sigma(u,a/L) &=& r_1^{(1)}\,  
   \Sigma^2\,  \frac{a}{2L} \, + \rmO(\Sigma^3), \label{e:deltact}
\end{eqnarray}
with $r_1^{(1)} = -0.013$ in the relevant range of $L/a \geq 8$. 
We have normalized by the one-loop 
contribution to $\ct$, using the known $\ct^{(1)}$.
In this way, $\Delta^{\ct}\Sigma$ gives the effect in $\Sigma$ 
if one takes as an uncertainty the one-loop term in the perturbative 
series of $\ct$. As here the one-loop term 
is the last known one, this is exactly what we want to do in
this work. 

As a check on the use of perturbation theory,
we performed simulations on our smallest lattice $L/a=8$ at $\bar g^2 \sim
4.5$ with three different values of $\ct$ around the one-loop one.
We found that the effective coefficient
\begin{equation}
        r_1^{\rm eff} \equiv \frac{2L}{a} \Sigma^{-2} \frac{\partial \Sigma}{\partial \ct} \,,                
\end{equation}
evaluates to
\begin{equation}
r_1^{\rm eff} = -0.0121(5) \frac{ g_0^2 \ct^{(1)}}{\Sigma^2}, 
\end{equation}
when we estimate it from a numerical derivative at our central simulation point $\ct=1+\ct^{(1)}g_0^2$.
The agreement with lowest order perturbation theory is 
good enough to just take \eq{e:deltact} as our estimate
of the uncertainty. 

We propagate (by quadrature) the full one-loop effect of this
boundary counterterm eq.~\eqref{e:deltact} to our error on
$\Sigma(u,a/L)$. Note that this effect is sub-dominant
in comparison with our statistical accuracy. 
The corresponding uncertainty due to $\cttil$ 
will be
neglected since it is suppressed
by a further power of $g^2$.


%% file: step.tex
As already mentioned, the way to connect non-perturbatively the
hadronic scale $L_{\rm had}$ and the intermediate scale $L_0$ passes
through the computation of the step scaling function. It is defined as
the continuum limit 
\begin{equation}
  \sigma(u) =  \lim_{a/L\rightarrow 0} \Sigma(u,a/L)\,,
\end{equation}
of its lattice approximation,
\begin{equation}
  \Sigma(u,a/L) = \gbar^2_{\rm GF}(2L)
  \Big|_{\gbar^2_{\rm GF}(L)=u, m=0}\,.
\end{equation}
The condition $m=0$ fixes the bare quark mass
for each resolution $a/L$ and each value of the bare coupling
$g_0^2$. The resulting function is denoted $m_{\rm cr}(g_0,a/L)$ and
described in appendix~\ref{ap:simulation}. The second condition,
$\gbar^2_{\rm GF}(L)=u$ fixes $g_0$ for each value of $u$
and resolution $a/L$ considered. The doubled lattices, where
$\gbar^2_{\rm GF}(2L)$ is determined, share the bare parameters with the $L/a$ lattices.

\subsection{Strategy and data set}

 In practice these conditions
have to be implemented by a tuning of the bare parameters such that the renormalized ones 
are fixed as described. We briefly explain our strategy to 
arrive at a precise tuning for a few appropriate values of $u$
and the estimates of $\Sigma$.

\input{tables/Sigma.tex}

\begin{enumerate}
\item 
The tuning of the bare mass $m_0$ was already carried out in
\cite{Nf3tuning} for the full range of bare couplings
and $a/L$ considered. In the continuum limit the chiral point
of vanishing quark mass is unique; the $a/L$-dependence 
is a cutoff effect. However, in order to have a smooth extrapolation
to the continuum limit, one first defines exactly which mass is
set to zero at a fixed $a/L$ and then determines the function
$m_{\rm cr}(g_0,a/L)$. In the cited reference this task was carried out
with high precision. As a result we can neglect any deviations from the 
exact critical line. The used functions $m_{\rm cr}(g_0,a/L)$ are listed in
\App{ap:simulation}.
\item
As a next step we performed 9 precise simulations with $L/a=16$. These
determine 9 values of $u=v_i, \; i=1,\ldots,9$, which we take as our
prime targets to compute $\sigma(\utarg_i)$. 
We further need values of $\beta$ for $L/a=8,12$ such that
  $\gbar_{\rm GF}^2$ equals our target values $\utarg_i$. This is achieved by
  an interpolation of several simulations described in detail in
  appendix~\ref{ap:tuning_gbar}. 
At this point we found for each $L/a=8,12,16$ nine values of $\beta$ where 
couplings $\gbar_{\rm GF}^2(L)$ match rather well. These $\beta$-values are listed in
\tab{tab:Sigma}.
\item 
We then carried out simulations on the doubled lattices
at the same values of $\beta,m_0$, see columns 4-6 in \tab{tab:Sigma}.
The data for $\gbar_{\rm GF}^2(2L)$ in the table are estimates of the 
step scaling function $\Sigma(u,a/L)$ at $u=\gbar_{\rm GF}^2(L)_{\beta,L/a}$.
As our estimates for $\gbar_{\rm GF}^2(L)$ we could take the numbers
from the interpolation in step 2. 
These are simply the same as those
at $L/a=16$. However, in order to enhance 
the precision, we perform separately at each $L/a=8,12$ 
an interpolating fit to all available data of \tab{tab:smallL}.
These fits determine $\gbar_{\rm GF}^2(L)$ in \tab{tab:Sigma}. 
Details on the very well determined interpolation are given
in  \App{ap:tuning_gbar}.
\item
As a last step we propagate the errors of $\gbar_{\rm GF}^2(L)$ into
those of $\Sigma$. As we will see in section~\ref{sec:ubyu} our
non-perturbative data is well described by the functional form 
\begin{equation}
  \frac{1}{\Sigma} - \frac{1}{u} = {\rm constant}\,, \label{e:shift}
\end{equation}
which suggests to use the derivative, $\partial \Sigma / \partial u =
\Sigma^2/u^2$ for the error propagation. 
This yields the last column of \tab{tab:Sigma}, where 
$u$ is the central value of $\gbar_{\rm GF}^2$ without error. 
The difference of the errors in columns 4 and 7 is 
mostly due to the uncertainty of $\rmO(a)$ improvement,
\eq{e:deltact}; a small part of the uncertainty is also
contributed by the propagated errors of $\gbar_{\rm GF}^2(L)$.
\end{enumerate}

The last two rows in \tab{tab:Sigma} are from additional simulations
performed with the aim of having $\gbar_{\rm GF}^2(2L)\approx 11.3$. They
will also be useful below.

\subsection{Continuum extrapolation of the step scaling function}
\label{s:fits}
The results at finite resolution need to be extrapolated to
the continuum. It is apparent from \tab{tab:Sigma} that this is an
essential step, since $\Sigma$ changes by up to 20\% 
in the accessible range of $a/L$ -- far outside the statistical errors.
However, our investigation in \sect{sec:cont} showed that 
the cutoff effects are strongly dominated by the $(a/L)^2$ 
terms, which motivates extrapolations linear in this variable.

Given the high precision which we achieve, this is a crucial part of
this work, and a detailed analysis will follow.
In particular, we first study 
the systematic effects in the continuum determination of $\sigma(u)$
by performing independent extrapolations at 9 fixed values of
$u$. These can transparently be illustrated by simple
graphs. 

\subsubsection{$\sigma(u)$ and systematic effects in
  the continuum extrapolations}
\label{sec:sys}
\label{sec:ubyu}
\begin{figure}
  \centering
  \includegraphics[width=0.9\textwidth]{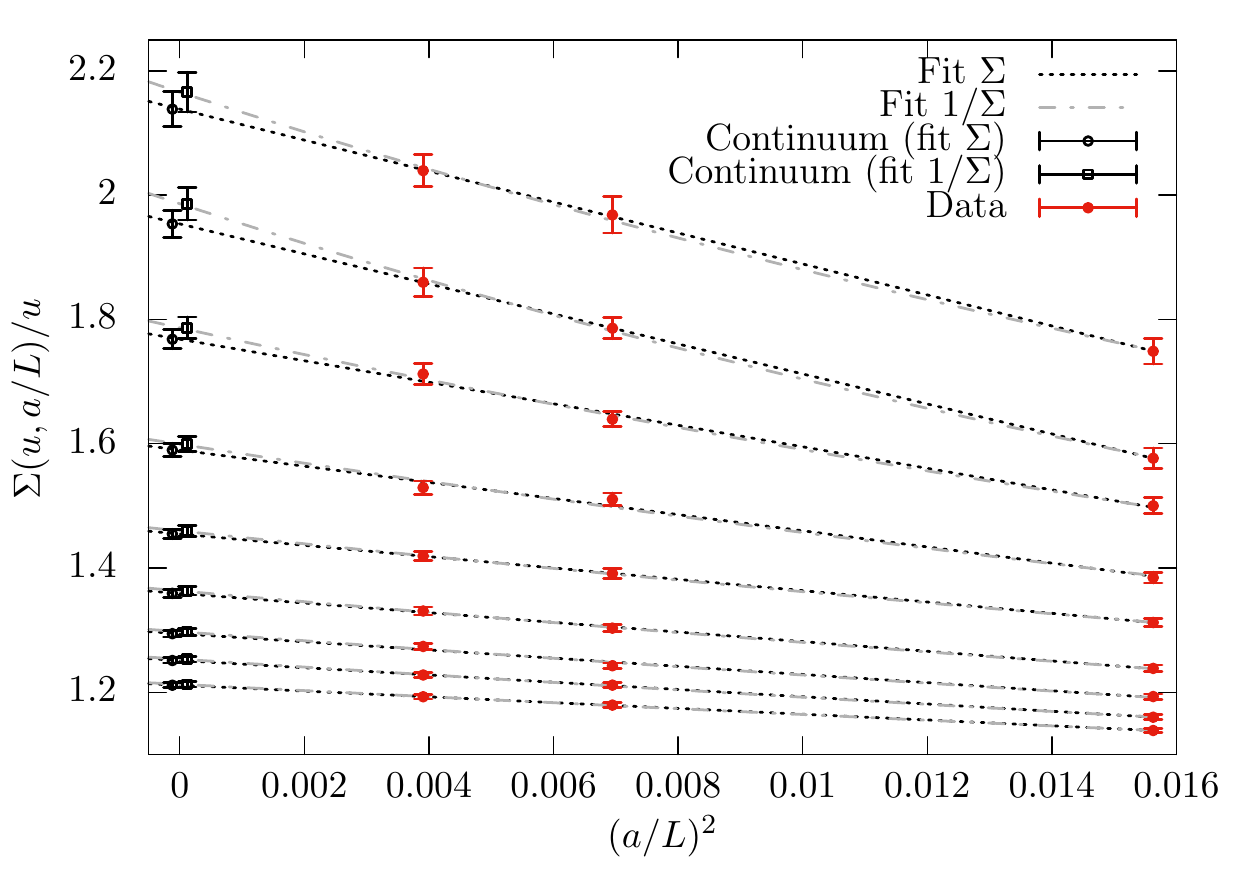}
  \caption{Continuum extrapolation of $\Sigma$ of data shifted
    to 9 different values of $u$. }
  \label{f:Sigma}
\end{figure}

Apart from the last two rows of \tab{tab:Sigma}, the deviations
of $\gbar_{\rm GF}^2(L)$ from the 9 target values $\utarg_i$ (the ones at $L/a=16$)
are very small. We can therefore simply shift the data for
$\Sigma$ using \eq{e:shift}.
The resulting data is shown in 
\fig{f:Sigma}. Within the uncertainties, linearity in $a^2$ is 
perfect and we extrapolate by 
\begin{eqnarray}
    \Sigma(\utarg_i,a/L) &=& \sigma_i + \tilde r_i \times (a/L)^2 \, ,
    \label{e:fitGF2a}
\end{eqnarray}
at each value $\utarg_i$. The quality of the fits is very good with
a total $\chi^2$ of 6.3 with 9 degrees of freedom.
The fit parameters $\sigma_i$, second column of \tab{t:sigmai}, 
are first estimates of the continuum step scaling function. 
It turns out that the non-perturbative results are well described by 
$1/\sigma_i - 1/\utarg_i \approx -0.083$ (see last two columns
of~\tab{t:sigmai}), which is the functional  
form of one-loop perturbation theory, but with a coefficient
slightly different from the perturbative $-0.0790$.
This surprising behavior holds out to $\sigma(u) = \rmO(10)$.
We will come to a comparison with perturbation theory 
later. For now this suggests to fit also
\begin{eqnarray}
    1/\Sigma(\utarg_i,a/L) &=& 1/\sigma_i + r_i \times (a/L)^2 \,.
    \label{e:fitGF2b}
\end{eqnarray}
The quality of these fits is as good as the previous ones ($\chi^2 =
6.3$ for 9 degrees of freedom). Discriminating statistically 
between the two fit forms would require far higher precision than 
we have.

\begin{figure}[tb]
     \centering
     \includegraphics[width=0.75\textwidth]{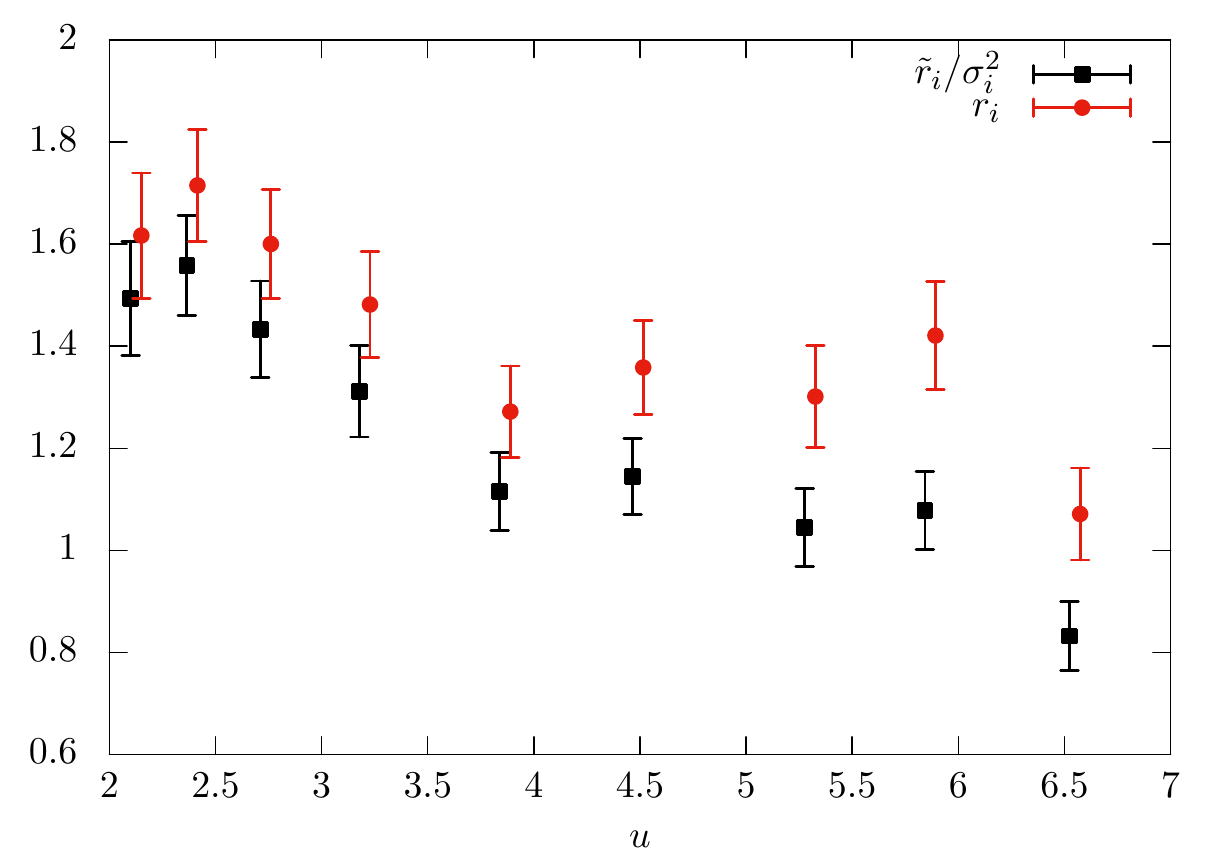}
     \caption{Slopes $r_i$ of \eq{e:fitGF2b} and 
     the in leading order in $a^2$ equivalent $-\tilde r_i/\sigma_i^2$ 
     with $\tilde r_i$ of \eq{e:fitGF2b}. 
   }
     \label{f:rhoi}
\end{figure}

An implicit assumption behind \eq{e:fitGF2a} and \eq{e:fitGF2b}
is that higher orders in $a^2$ are negligible. When this is the case,
the fit-parameters $\sigma_i$ have to agree between the two fits
(see \tab{t:sigmai}).
There is agreement at the level of one standard deviation.
However, the difference between the two extrapolations is of course systematic:
$\sigma_i$ are always larger when they are extrapolated following
\eq{e:fitGF2b}. This is also apparent in \fig{f:Sigma}. 
Furthermore, when nonlinearities in $a^2$ are negligible, there is
the more stringent condition $r_i = - \tilde r_i/\sigma_i^2$.
As expected, we find more significant differences between
these slope 
parameters\footnote{Note that 
the determination of asymptotic values of $r_i$ or $\tilde r_i$
is not our goal. We only discuss 
them because they show that differences
between the continuum limits estimated from 
\eq{e:fitGF2a} and \eq{e:fitGF2b} have to be taken into account.} (see
\fig{f:rhoi}). Note  
that the difference between the functional forms of \eq{e:fitGF2a}
and \eq{e:fitGF2b} is of order $a^4$. Due to
the relatively large $\rmO(a^2)$ effects, these are not negligible at 
large values of the coupling (at small values of $u$
we have good agreement between both type of fits). It is this ${\rmO}(a^4)$ 
effect that produces a  systematic shift in the parameters
$\sigma_i, r_i$.

\begin{table}[t!]
  \small
  \centering
  \begin{tabular}{lllll}\toprule
  $\utarg_i$ &\multicolumn{2}{c}{ $\sigma_i$ } 
    &\multicolumn{2}{c}{ $(1/\sigma_i - 1/\utarg_i)\times 10^{2}$ }  \\
        & \protect\eq{e:fitGF2a} & \protect\eq{e:fitGF2b} 
        & \protect\eq{e:fitGF2a} & \protect\eq{e:fitGF2b} \\ \midrule
    6.5489  & $   14.005(175)$        & $   14.184(197)$      & $-8.13(10)$  & $ -8.22(12)$ \\
    5.8673  & $   11.464(123)$        & $   11.654(146)$      & $-8.32(10)$  & $ -8.46(13)$ \\
    5.3013  & $\pz 9.371(\pz 79)$     & $\pz 9.468(\pz 89)$   & $-8.19(11)$  & $ -8.30(12)$ \\
    4.4901  & $\pz 7.139(\pz 47)$     & $\pz 7.181(\pz 51)$   & $-8.26(11)$  & $ -8.34(12)$ \\
    3.8643  & $\pz 5.622(\pz 28)$     & $\pz 5.641(\pz 30)$   & $-8.09(10)$  & $ -8.15(14)$ \\
    3.2029  & $\pz 4.354(\pz 19)$     & $\pz 4.367(\pz 21)$   & $-8.25(12)$  & $ -8.32(13)$ \\
    2.7359  & $\pz 3.541(\pz 14)$     & $\pz 3.550(\pz 15)$   & $-8.31(12)$  & $ -8.38(13)$ \\
    2.3900  & $\pz 2.991(\pz 10)$     & $\pz 2.996(\pz 10)$   & $-8.40(12)$  & $ -8.46(13)$ \\
    2.1257  & $\pz 2.575(\pz\pz 9)$   & $\pz 2.578(\pz\pz 9)$ & $-8.21(14)$  & $ -8.26(14)$ \\ \midrule
    \multicolumn{3}{l}{Constant fit:}                         & $-8.233(37)$ & $-8.316(42)$ \\
  \bottomrule
  \end{tabular}
  \caption{Examples for the continuum limits of the step scaling function\\ 
   $\sigma_i=\lim_{a/L\to0}\Sigma(v_i,a/L)$ obtained by various
   extrapolations at fixed values of $u=v_i$. The last row shows fits of
   columns 4 and 5 to a constant. These fits to a constant provide an
   excellent description of our data. 
}
\label{t:sigmai}
\end{table}

A fit of  $1/\sigma(u) - 1/u$ to a constant provides
a good description of our continuum data ($\chi^2/{\rm dof}< 1$) in the
whole range $u\in[2.1\,,\,6.5]$. Although the systematic difference
between the continuum fits \eq{e:fitGF2a} and \eq{e:fitGF2b} was 
point by point in $\sigma_i$ below our statistical accuracy, the uncertainty in a constant fit to $1/\sigma(u) - 1/u$
is reduced by a factor 3 due to the fact that we use 9 independent
values to determine it. The systematic effect then becomes clearly
noticeable.

\subsubsection{Fitting strategy}
\label{sec:sys2}

The previous considerations illustrate that the ${\rmO}(a^4)$
effects are not large, but still cannot simply be ignored. 
The size of the $\rmO(a^2)$
term, that 
amounts to 20\% at the largest value of the coupling $u_{\rm
  max}=6.5$ at $L/a=8$, suggests that there the $\rmO(a^4)$ effects are
around 5\%. Taking into account that a $u$-independent term is
removed by the normalization of the coupling,
this translates into the rough scaling
\begin{equation}
  \Delta^{\rm sys} \Sigma_i = 0.05\,\Sigma_i\, 
  \bigg(8\frac{a}{L}\bigg)^4\, \frac{u}{u_{\rm
        max}}\,. 
\end{equation}
This systematic effect is negligible compared with our
statistical accuracy for the
lattices with $L/a\ge 12$ at all values of $u$ (in fact the
differences seen in \tab{t:sigmai} become
insignificant when we perform the extrapolations
with just $L/a \ge 12$), but
it becomes dominant at $L/a=8$ and large values of $u$. 

When fitting to some particular functional form one performs
 a minimization of a $\chi^2$ function, defined as
\begin{equation}
  \chi^2(p_\alpha) = \sum_{i=1}^{N_{\rm data}} W_i\,\big[f(x_i;p_\alpha) -
    y_i\big]^2\,, \label{e:chisq}
\end{equation}
where $p_\alpha$ represent the parameters that describe the function
$f(x_i;p_{\alpha})$, and $x_i,y_i$ are the independent
and dependent variables, respectively. The weight, 
$W_i$, of each data point,
is usually taken from their uncertainty, but here
we should take into account that we cannot expect our data to be more accurately described
by a linear function
in $a^2$ than $\Delta^{\rm sys}\Sigma_i$. 
For the following we therefore define the weights by 
\begin{equation}
  \label{eq:fitw}
  W_i^{-1} = (\Delta\Sigma_i)^2 + (\Delta^{\rm sys}\Sigma_i)^2 \,,
\end{equation}
which strongly reduces the weights of the points further away 
from the continuum.  Note that we distinguish the 
weights of the fits from 
the errors $\Delta\Sigma_i$ of the data (statistical and the one
due to the uncertainty in $\ct$), which enter the error propagation
from the data to the parameters of the fit.

As an example for the consequences of introducing $W_i$, we repeat fits \eq{e:fitGF2a} and \eq{e:fitGF2b}.
We obtain continuum values $1/\sigma(u)-1/u$ which are still
perfectly described by a constant, but now the values of the
constants are $-0.0824(5)$ and $-0.0830(6)$,
respectively. Comparing with the last row of \tab{t:sigmai} we see
that uncertainties have increased and central values
are closer. Now both types of fits agree within one standard
deviation.

\subsubsection{Determination of $\sigma(u)$}

\begin{figure}[tb]
  \centering
  \includegraphics[width=\textwidth]{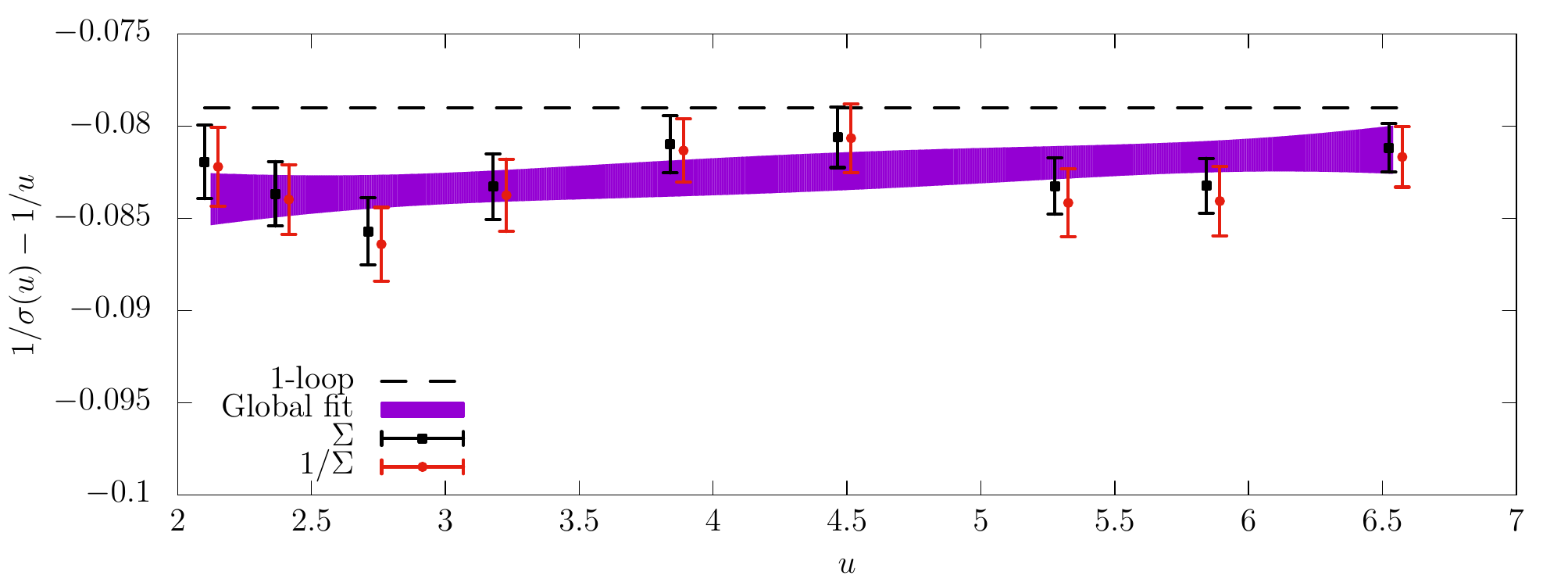}
  \caption{Comparison between different determinations of the
    continuum step scaling function $\sigma(u)$. Continuum
    extrapolations at fixed values of $u$ as described in \eq{e:fitGF2a} and
    \eq{e:fitGF2b} are labeled $\Sigma$ and $1/\Sigma$ respectively. A
    global fit with $\nsig=n_\rho=2$ is also shown,
    cf. \eq{eq:global}. The figure shows good agreement. 
}
  \label{f:global}
\end{figure}

As already noted, our non-perturbative data is very well described by
an effective one-loop functional form. This suggests two strategies
to determine the continuum step scaling function. First we can
perform continuum extrapolations at constant values of $u$ as
suggested in the previous sections (\eq{e:fitGF2a} and
\eq{e:fitGF2b}). The continuum values of $\sigma(\utarg_i)$ can then be fitted
to a functional form 
\begin{equation}
  \frac{1}{\sigma(\utarg_i)} - \frac{1}{\utarg_i} = Q(\utarg_i)\,,
  \qquad 
  Q(u)= \sum_{k=0}^{\nsig-1} c_k u^k\,.
  \label{e:Qfit}
\end{equation}
The number of parameters $\nsig$ is varied 
in order to check the stability of the procedure. Second, one
can also consider the 
possibility of combining the ansatz for the cutoff effects 
immediately with the parametrization of the continuum function
$\sigma(u)$ 
\begin{equation}
  \label{eq:global}
  \frac{1}{\Sigma(u,a/L)} - \frac{1}{u} = Q(u)
  + \rho(u) (a/L)^2\,.
\end{equation}
Apart from checking the stability of the procedure,
advantages of this global fit are as follows. The shifts to common values of $u$ for
different $a/L$ are not needed and the data in the last two rows  
of \tab{tab:Sigma} are easily included. Also more general forms of 
cutoff effects can be tried. Our investigation suggests that
\begin{eqnarray}
   \rho(u)  =\sum_{i=0}^{n_\rho-1} \rho_i u^i\,,
   \label{e:rho}
\end{eqnarray}
is a good parametrization of $\rho$ when at least 
$n_\rho =2$ terms are included. 

Figure~\ref{f:global} shows a comparison between the individual
extrapolations at fixed $u$ according to \eq{e:fitGF2a} and
\eq{e:fitGF2b}, and a global fit \eq{eq:global} with
$\nsig=n_\rho=2$. We recall that 
all fits are performed with the weights of \eq{eq:fitw}. 

\begin{table}[t!]
  \small
  \centering
  \begin{tabular}{cccc|llllll}
    \toprule
    Fit& $\nsig$& $n_\rho$& $W_i$ & $u_1$&$u_2$&$u_3$& $u_4$ & $s(g_1^2,g_2^2)$\\
    \midrule
      $\Sigma$,$\sigma$ &3& --  & $\Delta \Sigma_i^{-2}$ & $        5.866(21)$  & $        3.955(17)$  & $        2.981(13)$  & $        2.392(11)$  & -- 
  \\  $\Sigma$, $Q$ &3&--& $\Delta \Sigma_i^{-2}$   & $        5.867(21)$  & $        3.956(16)$  & $        2.981(14)$  & $        2.391(12)$  & --
  \\  $1/\Sigma$, $Q$ &3&--& $\Delta \Sigma_i^{-2}$   & $        5.832(21)$  & $        3.927(17)$  & $        2.960(13)$  & $        2.374(11)$  & -- 
  \\  $1/\Sigma$, $P$ &2& -- & $\Delta \Sigma_i^{-2}$ &  $        5.832(21)$  & $        3.927(15)$  & $        2.959(13)$  & $        2.374(11)$  & $       10.82(14)$ 
  \\  $1/\Sigma$, $P$ &3& -- & $\Delta \Sigma_i^{-2}$ &  $        5.831(21)$  & $        3.926(17)$  & $        2.959(13)$  & $        2.374(11)$  & $       10.82(15)$ 
  \\
  \midrule 
  $\Sigma$, $P$ &3& -- & (\ref{eq:fitw}) &  $        5.870(28)$  & $        3.954(22)$  & $        2.976(17)$  & $        2.385(15)$  & $       11.00(20)$ 
  \\ $1/\Sigma$, $P$ &1&3  & (\ref{eq:fitw}) & $        5.843(20)$  & $        3.939(18)$  & $        2.971(16)$  & $        2.385(13)$  & $       10.96(18)$ 
  \\  $1/\Sigma$, $P$ &2&3  & (\ref{eq:fitw}) & $        5.864(26)$  & $        3.944(19)$  & $        2.968(16)$  & $        2.378(14)$  & $       10.90(18)$ 
  \\  $1/\Sigma$, $P$ &3&3  & (\ref{eq:fitw}) & $        5.864(27)$  & $        3.944(21)$  & $        2.968(17)$  & $        2.378(14)$  & $       10.90(19)$ 
  \\ \eqref{eq:global2}, $P$ &2&2& (\ref{eq:fitw})   & $        5.872(27)$  & $        3.949(19)$  & $        2.971(16)$  & $        2.379(14)$  & $       10.93(19)$ 
  \\ \eqref{eq:global2}, $P$ &3&3& (\ref{eq:fitw})   & $        5.874(28)$  & $        3.951(22)$  & $        2.972(17)$  & $        2.379(14)$  & $       10.93(20)$ 
    \\
  \bottomrule 
  \end{tabular}
  \caption{Coupling sequence \eq{e:recurs} with $u_0=11.31$ 
  and scale factors $s(g_1^2,g_2^2)$ for $g_1^2=2.6723,\,g_2^2=11.31$
  for different fits to cutoff effects and the continuum 
  $\beta$-function. Fits are labelled by $\Sigma$ or $1/\Sigma$ 
  for continuum extrapolations according to \eq{e:fitGF2a} or \eq{e:fitGF2b} 
  respectively while the parametrization of the continuum step scaling
  function is labelled as $\sigma$ for 
  $\sigma(u)=u+s_0u^2 + s_1u^3 + u^3\sum_{n=1}^\nsig c_n u^n$ and labelled as
  $Q$ for \eq{e:Qfit}. Fits to the $\beta$-function (\eq{e:betafct}) are
  labelled $P$. For global fits
  we specify $n_\rho$, of \eq{e:rho}, while its absence indicates a
  fit of data extrapolated to the continuum at each value of $u=v_i$. 
  The weights $W_i$ refer to the definition of $\chi^2$, \eq{e:chisq}. 
  }
  \label{tab:ui}
\end{table}

A more quantitative test of the agreement between the 
$\sigma$ obtained from different
analysis is through the sequence $u_0=11.31,\, u_{i\geq1}$,
\eq{e:recurs}. We collect
this information in \tab{tab:ui}. Once the polynomial
is not too restricted, the results depend very
little on the number of terms $\nsig$ since we use this polynomial
interpolation only in the range where data are available.

\subsection{Determination of the $\beta$-function}

  \begin{figure}[t]
    \centering
    \includegraphics[width=0.9\textwidth]{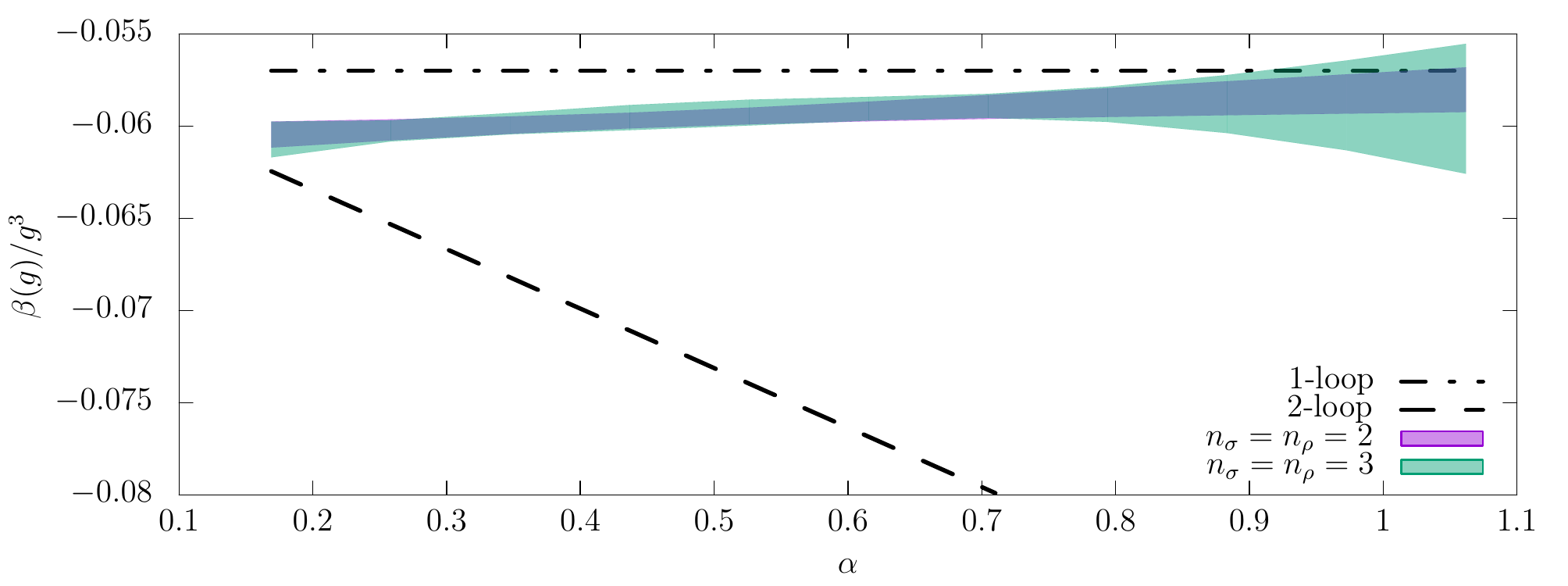}
    \caption{Comparison between two different fits to the
      $\beta$-function. They corresponds to the results quoted in the
      last two rows of table~\ref{tab:ui}.}
    \label{fig:beff}
  \end{figure}

Since our main goal is the determination of the scale factor $L_{\rm
  had}/L_0$ (see \eq{e:scalefactor}) it is very convenient
to replace the parametrization of $\sigma(u)$ by a 
parametrization of the $\beta$-function. Namely, we write
\begin{eqnarray}
  \beta(g) = - \frac{g^3}{P(g^2) }\,,\quad P(g^2) = p_0 +p_1 g^2 + p_2 g^4 + \ldots\,.
  \label{e:betafct}
\end{eqnarray}
The one-loop effective $\beta$-function just corresponds to the choice
$P(u) = p_0$, while higher order terms parameterize possible
(obviously small) deviations useful for a more detailed analysis and
an estimate of uncertainties. 
The step scaling function is then given by
\begin{eqnarray}
  \label{e:sigmaint}
  \begin{split}
    \log(2) &= 
    -\int_{\sqrt{u}}^{\sqrt{\sigma(u)}} \frac{\rmd x}{\beta(x)} = 
    \int_{\sqrt{u}}^{\sqrt{\sigma(u)}}\rmd x \frac{P(x^2)
    }{x^3}\\
    &= -\frac{p_0}{2}\left[\frac{1}{\sigma(u)} - \frac{1}{u}\right]
    + \frac{p_1}{2} \log\left[\frac{\sigma(u)}{u}\right]
    + \sum_{n=1}^{n_\mathrm{max}} \frac{p_{n+1}}{2n} 
    \left[\sigma^{n}(u) - u^{n}\right]\,,
  \end{split}
\end{eqnarray}
where $\nsig$ parameters correspond to 
$n_\mathrm{max} = \nsig-2$.
The parameters $p_i,\,i=0,\ldots, \nsig-1$ in \eq{e:betafct} can be obtained by fitting
our data for $\sigma(u)$ to \eq{e:sigmaint}. Any of our previous methods to
extrapolate the lattice step scaling function $\Sigma(u,a/L)$ to the
continuum can be used. In the case of the global fits, we
make use of a further variant
to parametrize the cutoff effects by fitting
\begin{equation}
  \label{eq:global2}
  \log(2) + \widetilde \rho(u)(a/L)^2 = 
  -\int_{\sqrt{u}}^{\sqrt{\Sigma(u,a/L)}} \frac{\rmd x}{\beta(x)}\,.
\end{equation}
Note that this fit ansatz differs from other global fits only by terms
$\rmO(a^4)$.  
Comparing the different approaches provides an
additional check that these effects are under control (see
discussion in sections~\ref{sec:sys}, \ref{sec:sys2}). 

Solving
numerically \eq{e:sigmaint} for $u$ we then compute the series of
couplings $u_i$. In table~\ref{tab:ui} we compare the results 
to those obtained
via the parameterizations of the step scaling function. There is good agreement between
different types of fits. 

Figure~\ref{fig:beff} shows a
comparison of the $\beta$-function obtained with two different fits. Their agreement underlines
that all uncertainties have been taken care of and 
that the small difference to the one-loop $\beta$-function 
is significant. At couplings $g^2\sim 3$ and larger, including the
universal two-loop term, $b_1 g^5$, 
in the $\beta$-function enlarges the difference. 
Therefore, perturbation theory is of little use in our range of couplings.

In the following we will use as our central result and
uncertainty the fit in the last row of the table. It has the largest
uncertainties and parameters
\begin{equation}
  p_0 = 16.26(69)\,,\quad p_1 = 0.12(26)\,,\quad p_2=-0.0038(211)\,,
\end{equation}
with covariance matrix 
\begin{equation}
  {\rm cov}(p_i,p_j) = \left(
    \begin{array}{rrr}
      4.78071\times 10^{-1}& -1.76116\times 10^{-1}& 1.35305\times 10^{-2}\\
      -1.76116\times 10^{-1}& 6.96489\times 10^{-2}& -5.54431\times 10^{-3}\\
      1.35305\times 10^{-2}& -5.54431\times 10^{-3}& 4.54180\times 10^{-4}\\
    \end{array}
\right)\,.
\end{equation}


%% file: tables/Sigma.tex
\begin{table}[p!]
  \small
  \centering
  \begin{tabular}{cccccccc}\toprule
   $L/a$ & $\beta$    & $\gbar^2(L)$    & $\gbar^2(2L)$         & $\Nms$      & $N_Q$           & $\Sigma(u,a/L)$ \\\midrule 
    $ 8$ & $3.556470$ & $6.5485(60)$    & $11.452(79)$          & $2000,2000$ & $725,\emptyset$ & $  11.452(134)$  \\  
    $ 8$ & $3.653850$ & $5.8670(34)$    & $\pz 9.250(66)$       & $2000$      & $220$           & $\pz 9.250(\pz 97)$  \\  
    $ 8$ & $3.754890$ & $5.3009(32)$    & $\pz 7.953(44)$       & $2001$      & $30 $           & $\pz 7.953(\pz 68)$  \\  
    $ 8$ & $3.947900$ & $4.4848(25)$    & $\pz 6.207(23)$       & $2001$      & $1  $           & $\pz 6.207(\pz 39)$  \\  
    $ 8$ & $4.151900$ & $3.8636(21)$    & $\pz 5.070(16)$       & $2001$      & $0  $           & $\pz 5.070(\pz 26)$  \\  
    $ 8$ & $4.457600$ & $3.2040(18)$    & $\pz 3.968(11)$       & $2001$      & $0  $           & $\pz 3.968(\pz 17)$  \\  
    $ 8$ & $4.764900$ & $2.7363(14)$    & $\pz 3.265(\pz 8)$    & $2001$      & $0  $           & $\pz 3.265(\pz 12)$  \\  
    $ 8$ & $5.071000$ & $2.3898(15)$    & $\pz 2.772(\pz 6)$    & $2001$      & $0  $           & $\pz 2.772(\pz \pz 9)$  \\  
    $ 8$ & $5.371500$ & $2.1275(15)$    & $\pz 2.423(\pz 5)$    & $2001$      & $0  $           & $\pz 2.423(\pz \pz 7)$  \\\cmidrule(lr){1-7}
    $12$ & $3.735394$ & $6.5442(82)$    & $12.874(165)$         & $3000$      & $\emptyset$     & $   12.874(191)$  \\  
    $12$ & $3.833254$ & $5.8728(46)$    & $10.497(\pz 78)$      & $2400$      & $\emptyset$     & $   10.497(\pz 99) $  \\  
    $12$ & $3.936816$ & $5.2990(36)$    & $\pz 8.686(\pz 49)$   & $2400$      & $\emptyset$     & $\pz 8.686(\pz 64)$  \\  
    $12$ & $4.128217$ & $4.4908(32)$    & $\pz 6.785(\pz 36)$   & $2400$      & $1$             & $\pz 6.785(\pz 44)$  \\  
    $12$ & $4.331660$ & $3.8666(25)$    & $\pz 5.380(\pz 25)$   & $2400$      & $0$             & $\pz 5.380(\pz 29)$  \\  
    $12$ & $4.634654$ & $3.2058(17)$    & $\pz 4.180(\pz 14)$   & $2403$      & $0$             & $\pz 4.180(\pz 17)$  \\  
    $12$ & $4.938726$ & $2.7380(15)$    & $\pz 3.403(\pz 11)$   & $2400$      & $0$             & $\pz 3.403(\pz 13)$  \\  
    $12$ & $5.242465$ & $2.3902(11)$    & $\pz 2.896(\pz \pz 9)$& $2400$      & $0$             & $\pz 2.896(\pz 10)$  \\  
    $12$ & $5.543070$ & $2.1235(12)$    & $\pz 2.504(\pz \pz 8)$& $2400$      & $0$             & $\pz 2.504(\pz \pz 9)$  \\\cmidrule(lr){1-7} 
    $16$ & $3.900000$ & $6.5489(155)$   & $   13.357(136)$      & $1205$      & $\emptyset$     & $   13.357(167)$  \\  
    $16$ & $4.000000$ & $5.8673(140)$   & $   10.913(118)$      & $1404$      & $\emptyset$     & $   10.913(136)$  \\  
    $16$ & $4.100000$ & $5.3013(134)$   & $\pz 9.077(\pz 75)$   & $1403$      & $1$             & $\pz 9.077(\pz 91)$  \\  
    $16$ & $4.300000$ & $4.4901(\pz 77)$& $\pz 6.868(\pz 40)$   & $2507$      & $0$             & $\pz 6.868(\pz 48)$  \\  
    $16$ & $4.500000$ & $3.8643(\pz 63)$& $\pz 5.485(\pz 22)$   & $2000$      & $0$             & $\pz 5.485(\pz 28)$  \\  
    $16$ & $4.800000$ & $3.2029(\pz 52)$& $\pz 4.263(\pz 16)$   & $2000$      & $0$             & $\pz 4.263(\pz 20)$  \\  
    $16$ & $5.100000$ & $2.7359(\pz 35)$& $\pz 3.485(\pz 11)$   & $2500$      & $0$             & $\pz 3.485(\pz 14)$  \\  
    $16$ & $5.400000$ & $2.3900(\pz 30)$& $\pz 2.935(\pz\pz 7)$ & $2500$      & $0$             & $\pz 2.935(\pz \pz 9)$  \\  
    $16$ & $5.700000$ & $2.1257(\pz 25)$& $\pz 2.536(\pz\pz 7)$ & $2500$      & $0$             & $\pz 2.536(\pz \pz 8)$  \\\cmidrule(lr){1-7}  
    $12$ & $3.793389$ & $6.1291(56)$    & $11.788(132)$         & $2556$      & $\emptyset$     & $  11.788(154) $  \\  
    $16$ & $3.976400$ & $6.037(14)\pz$  & $11.346(100)$         & $1203$      & $\emptyset$     & $  11.346(124)$  \\  
   \bottomrule
  \end{tabular}
  \caption{Step scaling functions. At the specified $\beta$, 
  we list $\gbar^2(L)$ on the $L/a$-lattice obtained from the described fit
  as well as $\gbar^2(2L)$ on the $2L/a$-lattice.  Their errors do not contain
  the uncertainty of $\ct$. $N_\mathrm{ms}$ and  $N_Q$ 
  refer to the measurements on the $2L/a$-lattice 
  Simulations with $N_Q=\emptyset$ were 
     carried out with the algorithm restricted to $Q=0$. At $\beta= 3.556470$
  and $L/a=16$ we have two ensembles, with and without fixing the
  topology (both ensembles give compatible results and in columns 4,7
  we quote as results the weighted avarage).
  The last column contains $\Sigma(u,a/L)$ with
  $u$ equal to the central value of column 3 and the full error obtained
  from $\gbar^2(L)$, $\gbar^2(2L)$ as well as the uncertainty of $\ct$.
  Note that errors in columns 3 and 7 are correlated, as 
  discussed in the text. 
  }
\label{tab:Sigma}
\end{table}



%% file: matchL0.tex
\subsection{Matching with the scale $1/L_0$}

In this section we relate the scale $1/L_0$ defined
in~\cite{Brida:2016flw} by the condition 
\begin{equation}
  \bar g_{\rm SF}^2(L_0) = 2.012\,,
\end{equation}
to the coupling in our GF scheme. More precisely,
we define the function
\begin{equation}
  \varphi(u) =  \lim_{a/L\rightarrow 0} \Phi(u,a/L)\,,
\end{equation}
with
\begin{equation}
  \Phi(u,a/L) = \gbar^2_{\rm GF}(2L)
  \Big|_{\gbar^2_{\rm SF}(L)=u, m=0}\,.
\end{equation}
Recall that the SF coupling is defined with a
background field, while the boundary conditions of our gradient
flow scheme correspond to a zero background field. The connection
between the couplings goes through the common bare parameters
defined by the condition $\gbar^2_{\rm SF}(L)=u, m=0$, together
with the resolution $a/L$.

\begin{figure}
  \centering
  \includegraphics[width=0.8\textwidth]{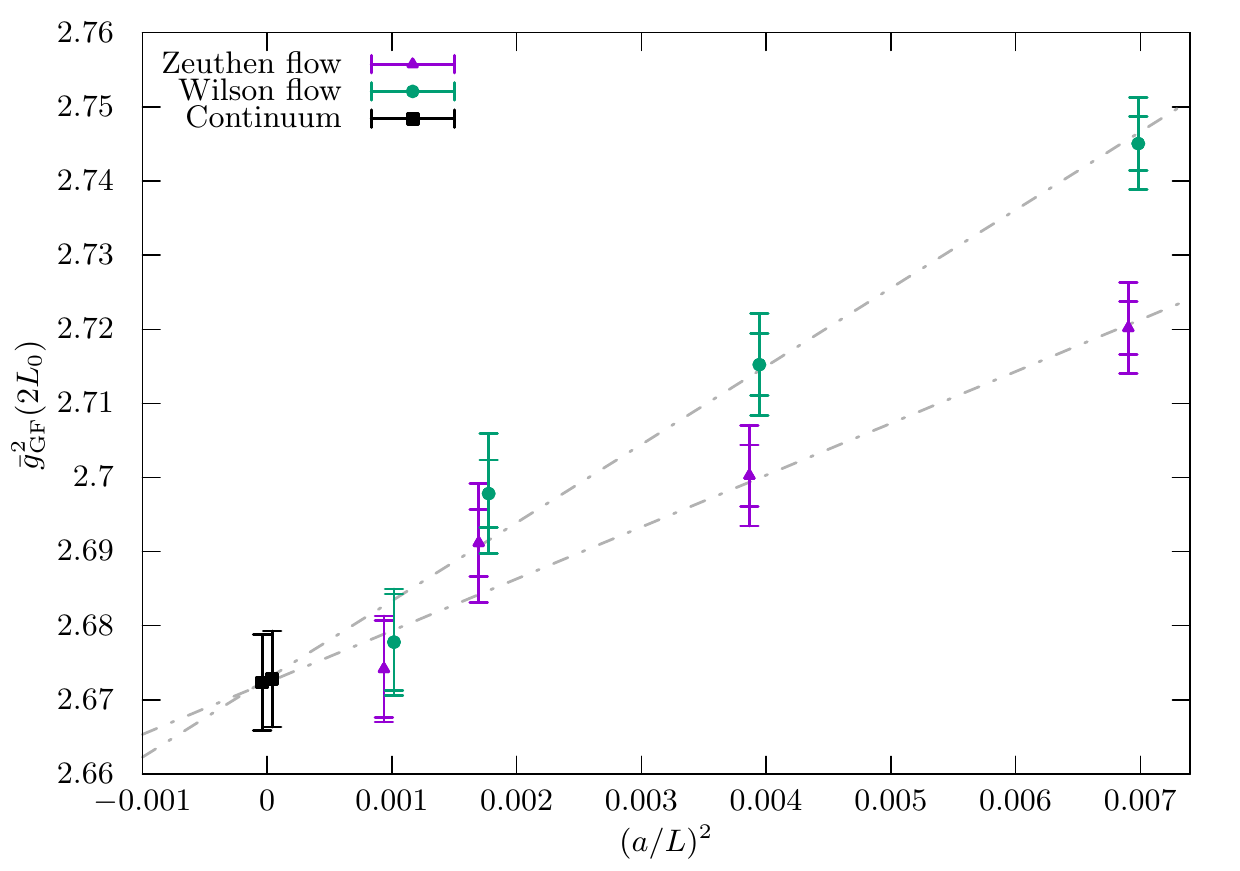}
  \caption{Continuum extrapolation of $\bar g^2_{\rm GF}(2L_0)$
    with the bare parameters determined by the condition
    $\bar g_{\rm SF}^2(L_0) = 2.012$. The continuum
    extrapolation is performed using both the Wilson flow/Clover
    discretization and our preferred setup Zeuthen flow/LW observable
    (the latter shows smaller discretization effects). The two types
    of error bars for each data point correspond to the inclusion or
    not of the propagated error for the SF coupling, cf.~text.}
  \label{fig:Lsw}
\end{figure}

We do not need the functional dependence on $u$, but rather just 
the single value $\varphi(2.012)$. We combine the change
of schemes SF $\to$ GF with a scale change by a factor of two,
because this avoids the disadvantages of both schemes 
at the same time: $\gbar^2_{\rm GF}$ has noticeable cutoff effects
when $a/L$ is too small and $\gbar^2_{\rm SF}$ needs very large
statistics if $L/a$ is too large. A last choice to make is
the discretization. Here we choose the Wilson gauge action where
the counter-terms (coefficients $\ct, \cttil$, see \cite{Brida:2016flw})  
which cancel linear $a$ effects 
are
perturbatively known, such that they are suppressed to
the negligible level of $g^8 a/L$. 
The action as well as the
definition of the critical line $m=0$ is exactly
as in \cite{Brida:2016flw,paperSF}. In fact, with the exception
of $L/a=16$, the numerical values of 
$\beta,\kappa,\gbar^2_\mathrm{SF}$
in \tab{t:switch} are taken from there, interpolated to the fixed 
value $\gbar_\mathrm{SF}^2=2.012$. More details will be given
elsewhere~\cite{paperSF}. Our measurements of the 
GF coupling on the doubled lattices (``Zeuthen flow'')
are listed in table~\ref{t:switch}. The errors in the
last column include the errors of $\bar g^2_{\rm SF}(L_0)$ (column 4 of
\tab{t:switch}). Like for the step-scaling function in \eq{e:shift}, we use
the derivative $\partial_u \Phi(u,a/L)\simeq
\Phi(u,a/L)^2/u^2$ for the Gaussian error propagation. The
additional error does not depend very much on this particular ansatz
and is subdominant, as can be 
also seen in table~\ref{t:switch} and in \fig{fig:Lsw} where the errors
both before and after error propagation are shown.

\begin{table}[t!]
  \small
  \centering
  \begin{tabular}{clccccc}
    \toprule
  $L/a$ & $\quad\beta$ & $\kappa$ & $\gbar_\mathrm{SF}^2(L)$ & 
  $\gbar_\mathrm{GF}^2(2L)$ &  $\Phi(u,a/L)$ &   \\
  \midrule
6& 6.2735  & 0.1355713 & 2.0120(27)&   2.7202(36)&  2.7202(61)&   \\
8& 6.4680  & 0.1352363 & 2.0120(30)&   2.7003(41)&  2.7003(68)&  \\
12&6.72995 & 0.1347582 & 2.0120(37)&   2.6912(45)&  2.6912(80)&   \\
16&6.9346  & 0.1344121 & 2.0120(17)&   2.6742(65)&  2.6742(72) \\
\midrule
\multicolumn{4}{l}{continuum limit} && 2.6723(64) \\
    \bottomrule
  \end{tabular}
  \caption{  Data for both the SF and GF couplings as required for the
    matching procedure.
}
\label{t:switch}
\end{table}

The continuum extrapolation of $\Phi$ can be seen in~\fig{fig:Lsw}.
We also show results with the Wilson flow, but
the Zeuthen flow
eq.~\eqref{eq:lattcoupling} has smaller cutoff effects. Due to the very high
statistical correlation of the numbers, a combination of 
the two discretizations of the flow observable does not lead to 
an improvement of the final errors. 
We therefore quote only the continuum limit from the Zeuthen flow.  
The main result of this section  is then
\begin{equation}
  \bar g_{\rm GF}^2(2L_0) = \varphi(2.012) = 2.6723(64)\,.
  \label{e:phi2}
\end{equation}

\subsection{The ratio $L_{\rm had} / L_0$ }

Using our fits to  the $\beta$-function, the 
scale factor $s=L_2/L_1$ between $g_2=\gbar(L_2)$ and $g_1=\gbar(L_1)$
can be easily computed via
\begin{eqnarray}
  \log(s(g_1^2,g_2^2)) = \int_{g_1}^{g_2}\rmd x \frac{P(x^2) }{x^3} 
  = \frac{p_0}{2g_1^2} -  \frac{p_0}{2g_2^2} + p_1 
  \log\left(\frac{g_2}{g_1}\right) 
  + \sum_{n=1}^{n_\mathrm{max}}
  \frac{p_{n+1}}{2n} [g_2^{2n} - g_1^{2n}] \,.
\end{eqnarray}
Numbers for $s$ from the various fits are shown in the last column of
\tab{tab:ui}. They refer to our default value $g_2^2=\gbar_{\rm
  GF}^2(L_{\rm had})=11.31$ defining 
$L_\mathrm{had}$ and $g_1^2=\gbar_{\rm GF}^2(2L_0)=2.6723$ given by
the central value of  
$\varphi(2.012)$ determined above. 
The error of $\varphi$
can be propagated straightforwardly, yielding
\begin{equation}
  L_\mathrm{had}/L_0 =  21.86(42)\,.
  \label{e:LhadL0}
\end{equation}
As $\Lambda_{\msbar}^{(3)} =  0.0791(21)/ L_0 $ is known~\cite{Brida:2016flw},
the last step  on the way to a determination of the 
$\Lambda$-parameter
in physical units is the computation of a physical observable
of dimension mass in large volume and at the physical masses of
the three quarks. This has to be combined with $L_\mathrm{had}/a$
at identical bare couplings and extrapolated to $a=0$. 
Passing this last milestone still needs input from the
\texttt{CLS} ensembles \cite{Bruno:2014jqa}.


%% file: conclusions.tex
\begin{figure}[t]
  \centering
  \includegraphics[width=\textwidth]{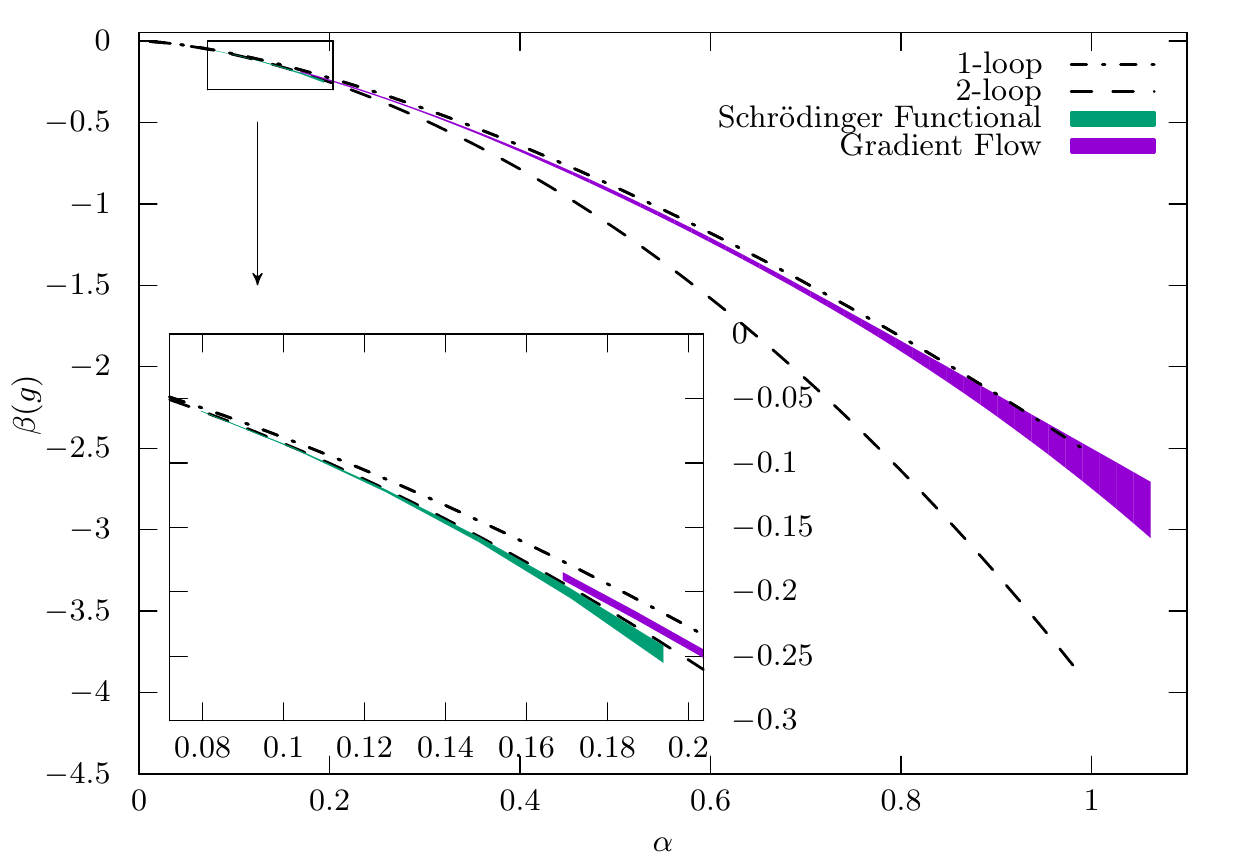}
  \caption{The $\beta$-functions in the SF-scheme from an effective
  4-loop fit in \cite{Brida:2016flw} and in the GF scheme evaluated
  in this work.}
  \label{f:beta}
\end{figure}

The main goal of this work was to connect the 
(technical) scales $L_0$ and $L_\mathrm{had}$ precisely. 
This is one of the three steps leading to a determination 
of the three-flavor $\Lambda$-parameter in physical units. 

The precision of the result, \eq{e:LhadL0}, is rather remarkable
since such a scale ratio can only be determined through the 
running of a coupling and a step scaling strategy
\cite{Luscher:1991wu} --- at least if one wants to obtain a purely
non-perturbative result  
and a controlled continuum limit.
Since couplings usually run relatively slowly it is necessary to determine
this running with extreme precision in order to achieve the 2\%
accuracy on the scale ratio. 
Through the gradient flow \cite{Luscher:2010iy} running coupling in a
finite volume\cite{Fritzsch:2013je} we achieved excellent precision.  
However, scaling violations had to be dealt with very carefully.  
After applying systematic Symanzik improvement \cite{Ramos:2015baa}, 
they were still very significant, but we could show that
they are rather accurately described by an $a^2$ behavior 
when the flow time, $t$, satisfies $a^2/(8t) < 0.3$. Since 
we chose our lattice spacings small enough, we could extrapolate 
to the continuum with three resolutions. All in all this milestone
on the way to a precise $\Lambda$-parameter has been passed.

Let us discuss also what else we have learned on the way.
The behavior of the step scaling function, \fig{f:global},
is rather surprising. It follows the one-loop functional form 
very precisely, but with a coefficient slightly different from the
universal perturbative one, out to large values of the coupling. For further
details, one better considers  
the $\beta$-function (\fig{fig:beff}). Here, the non-perturbative result is
in the middle between one-loop and two-loop at our smallest coupling, 
$\alpha=\gbar_{\rm GF}^2(L)/(4\pi)=0.17$.  Describing this
by higher order perturbation theory requires a large 
three-loop coefficient and therefore signals the breaking down
of perturbation theory at this coupling or close by. 
One might consider the statistical 
significance at the weakest coupling in \fig{fig:beff}
insufficient for a strong conclusion, but the effect 
becomes increasingly significant at
larger $\alpha$. For example at $\alpha=0.25$, still a coupling where 
perturbation theory is routinely used,  
the non-perturbative
running is many standard deviations away from two-loop.
Perturbation theory has broken down.
This finding reinforces what we saw before in the SF-schemes,
where the region below $\alpha\approx0.2$ was studied \cite{Brida:2016flw}.  The $\beta$-function in one of the schemes ($\nu=0$) discussed in
\cite{Brida:2016flw} is close to the known three-loop one, 
while other schemes are significantly off. 
In \fig{f:beta} we plot it together with the GF-scheme used
in this paper.  
For the GF-scheme we show only the range of couplings covered by our data. In contrast, for the SF-scheme, we show it all 
the way to $g=0$, since the connection to the asymptotic perturbative 
behavior was convincingly established.
The figure provides a warning that perturbation theory needs to
be applied with great care in the sense that its asymptotic nature 
should not be forgotten. For more details we refer to \cite{Brida:2016flw}. The figure also summarizes well where we stand 
concerning the determination of $\Lambda$. ``Only'' the 
very low energy connection of the GF-scheme to the hadronic world
remains to be carried through. The \texttt{CLS} simulations will allow us to
achieve this with an estimated $1 - 1.5\%$ precision 
\cite{Bruno:2014jqa,Bruno:2014lra,alphaprep}. 
For now, let us just mention that a combination of
the rough result $\gbar_{\rm GF}^2(L)=11$ for $\beta=3.55,\;L/a=16$ 
with the lattice spacing of \cite{Bruno:2014jqa} yields 
$L_\mathrm{had}=1$~fm. We have therefore computed the running 
in a range of $\mu$ from around 200~MeV to 4~GeV.


%% file: app_tuning.tex
\begin{table}[tb!]
  \small
  \centering
  \begin{tabular}{cccrccccc}\toprule
    $L/a$ & $\beta$     & $\bar g^2_{\rm GF}$ 
                                      & $r_a$    & $N_{\rm poles}$ 
                                                        & $\lambda_a^{\rm min}\times 10^{2}$
                                                                    & $\langle \lambda_a\rangle\times10^{2}$
                                                                             & $\langle \lambda_b\rangle$
                                                                                              & $\lambda_b^{\rm max}$ \\\midrule
     $24$ &  $3.735394$ & $   12.874$ & $0.010$  & $12$ & $1.44$ & $3.45(8)$ & $6.045(1)\pz $ & $6.62$ \\
     $24$ &  $3.793389$ & $   11.788$ & $0.010$  & $12$ & $1.89$ & $3.78(5)$ & $5.992(1)\pz $ & $6.26$ \\
     $24$ &  $3.833254$ & $   10.497$ & $0.015$  & $12$ & $2.33$ & $4.19(2)$ & $5.956(1)\pz $ & $6.19$ \\
     $24$ &  $3.936816$ & $\pz 8.686$ & $0.020$  & $11$ & $2.76$ & $4.69(3)$ & $5.8792(9)$    & $6.29$ \\
     $24$ &  $4.128217$ & $\pz 6.785$ & $0.025$  & $10$ & $3.46$ & $5.47(3)$ & $5.7648(8)$    & $6.01$ \\
     $24$ &  $4.331660$ & $\pz 5.380$ & $0.025$  & $10$ & $4.30$ & $6.01(3)$ & $5.6731(7)$    & $5.88$ \\
     $24$ &  $4.634654$ & $\pz 4.180$ & $0.025$  & $10$ & $5.24$ & $6.81(2)$ & $5.5739(7)$    & $5.77$ \\
     $24$ &  $4.938726$ & $\pz 3.403$ & $0.025$  & $10$ & $5.70$ & $7.39(2)$ & $5.5012(6)$    & $5.69$ \\
     $24$ &  $5.242465$ & $\pz 2.896$ & $0.025$  & $10$ & $6.29$ & $7.90(2)$ & $5.4457(6)$    & $5.64$ \\
     $24$ &  $5.543070$ & $\pz 2.504$ & $0.025$  & $10$ & $6.75$ & $8.28(1)$ & $5.4036(6)$    & $5.66$ \\\cmidrule(lr){1-9}
     $32$ &  $3.900000$ & $   13.357$ & $0.0075$ & $12$ & $1.15$ & $2.52(9)$ & $5.949(1)\pz $ & $6.20$ \\
     $32$ &  $3.976400$ & $   11.346$ & $0.0075$ & $12$ & $1.84$ & $2.96(4)$ & $5.895(1)\pz $ & $6.09$ \\
     $32$ &  $4.000000$ & $   10.913$ & $0.0100$ & $12$ & $1.84$ & $3.09(4)$ & $5.878(1)\pz $ & $6.07$ \\
     $32$ &  $4.100000$ & $\pz 9.077$ & $0.0100$ & $12$ & $2.02$ & $3.40(3)$ & $5.821(1)\pz $ & $6.05$ \\
     $32$ &  $4.300000$ & $\pz 6.868$ & $0.0100$ & $11$ & $2.78$ & $4.01(3)$ & $5.7213(8)$    & $5.94$ \\
     $32$ &  $4.500000$ & $\pz 5.485$ & $0.0100$ & $11$ & $3.16$ & $4.49(2)$ & $5.6495(7)$    & $5.83$ \\
     $32$ &  $4.800000$ & $\pz 4.263$ & $0.0100$ & $11$ & $3.79$ & $4.97(2)$ & $5.5645(7)$    & $5.74$ \\
     $32$ &  $5.100000$ & $\pz 3.485$ & $0.0100$ & $11$ & $4.31$ & $5.42(1)$ & $5.501(2)\pz $ & $5.67$ \\
     $32$ &  $5.400000$ & $\pz 2.935$ & $0.0100$ & $11$ & $4.86$ & $5.76(2)$ & $5.450(1)\pz $ & $5.57$ \\
     $32$ &  $5.700000$ & $\pz 2.536$ & $0.0100$ & $11$ & $5.22$ & $6.09(1)$ & $5.408(1)\pz $ & $5.56$ \\
     \bottomrule
  \end{tabular}
  \caption{Parameter $r_a$ determining the interval of the Zolotarev
    approximation (we always choose $r_b=7.5$),  
  and the corresponding number of poles, $N_{\rm poles}$. We also report the measured values 
  $\lambda_a,\lambda_b$ in our runs.}
  \label{tab:rarb}
\end{table}

\subsection{Algorithms, simulation parameters, and autocorrelations}

In this work we simulated with a modified version of the
\texttt{openQCD v1.0}  
package~\cite{Luscher:2012av}, using a Hasenbusch-type splitting of
the quark  
determinant for two of our mass-degenerate quarks~\cite{Hasenbusch:2001ne,Hasenbusch:2002ai}, and an RHMC~\cite{Kennedy:1998cu,Clark:2006fx}
for the third one. Apart from boundary terms, we have the same action as 
\texttt{CLS.} The interested reader may find it useful to
consult~\cite{Bruno:2014jqa}, 
where those simulations are described. Here we focus on some peculiarities
of our finite volume simulations: the projection to the zero
topological charge 
sector, the scaling of the spectral gap of the Dirac operator, and the
behavior 
of the integrated autocorrelation times of the renormalized
coupling. The latter 
characterize the performance of the algorithm and hence the effort
which we put 
into the computation.

\subsubsection{Algorithms}

An important speed up in HMC simulations is gained by splitting the contribution
of two of the quarks, $(\det D)^2=\det(D^\dagger D)$, into several factors~\cite{Hasenbusch:2001ne},
and representing each factor by a separate pseudo-fermion field. For our expensive 
simulations with $L/a=24,32$, we used three factors. More precisely, the splitting 
is characterized by the mass-parameters: $\mu_0=0$, $\mu_1=0.1$, and $\mu_2=1.2$, in the 
notation of~\cite{Luscher:2012av}. Having $\mu_0=0$ means that twisted mass reweighting, 
which is also implemented in the package, is not used. We find that this is not necessary, 
as the finite volume operator $D^\dagger D$ has a sufficiently stable gap. We shall show
results for the gap in \App{a:gap}.

A peculiar aspect of our finite volume renormalization scheme is that we are only interested
in expectation values obtained in the zero topological sector (see eq.~\eqref{eq:coupling}). 
On the lattice, this is implemented by using the definition of the topological charge at
positive flow time (see eqs.~(\ref{eq:qtop}--\ref{eq:deltaq}) and~\cite{Ce:2015qha} for 
more information). We explored two possibilities in order to obtain these expectation values:
\begin{description}
\item[\texttt{Algorithm A:}] 
Use a standard simulation and include the term $\hat \delta(Q)$, 
\eq{eq:deltaq}, as part of the definition of the observable. 

\item[\texttt{Algorithm B:}] Include the factor $\hat\delta(Q)$
  as part of the Boltzmann weight and generate an ensemble that only contains 
  configurations with $|Q|<0.5$. This is easily implemented by adding an accept/reject
  step after each trajectory. 
\end{description}
A consistency check between the two procedures was performed by generating two ensembles 
at $L/a=8$, $\beta= 3.556470$: one with \texttt{Algorithm A}, and a second with \texttt{Algorithm B}, 
obtaining respectively $\gbar_{\rm GF}^2(2L)=11.54(11)$  and
$\gbar_{\rm GF}^2(2L)=11.39(11)$. The average is given in
table~\ref{tab:Sigma}.  

In our tables we use $N_Q$ to denote the number of configurations that have $|Q|\ge 0.5$, 
and therefore do not contribute to the determination of expectation values. The symbol
$\emptyset$, instead, denotes an ensemble produced with \texttt{Algorithm B}. These 
ensembles have $|Q|<0.5$ throughout. A downside of \texttt{Algorithm B} is that
the acceptance rate can drop significantly below 1, with our lowest value being~$0.65$. 
This low acceptance rate is due to attempts of the algorithm to enter other topological 
sectors, and not to large violations of the HMC energy conservation. As this happens only
at the coarse lattice spacings, one can usually choose an efficient algorithm between \texttt{A} 
and \texttt{B} for a given choice of parameters; at least in the range we considered.

\begin{figure}
  \centering
  \begin{subfigure}[b]{0.48\textwidth}
    \includegraphics[width=\textwidth]{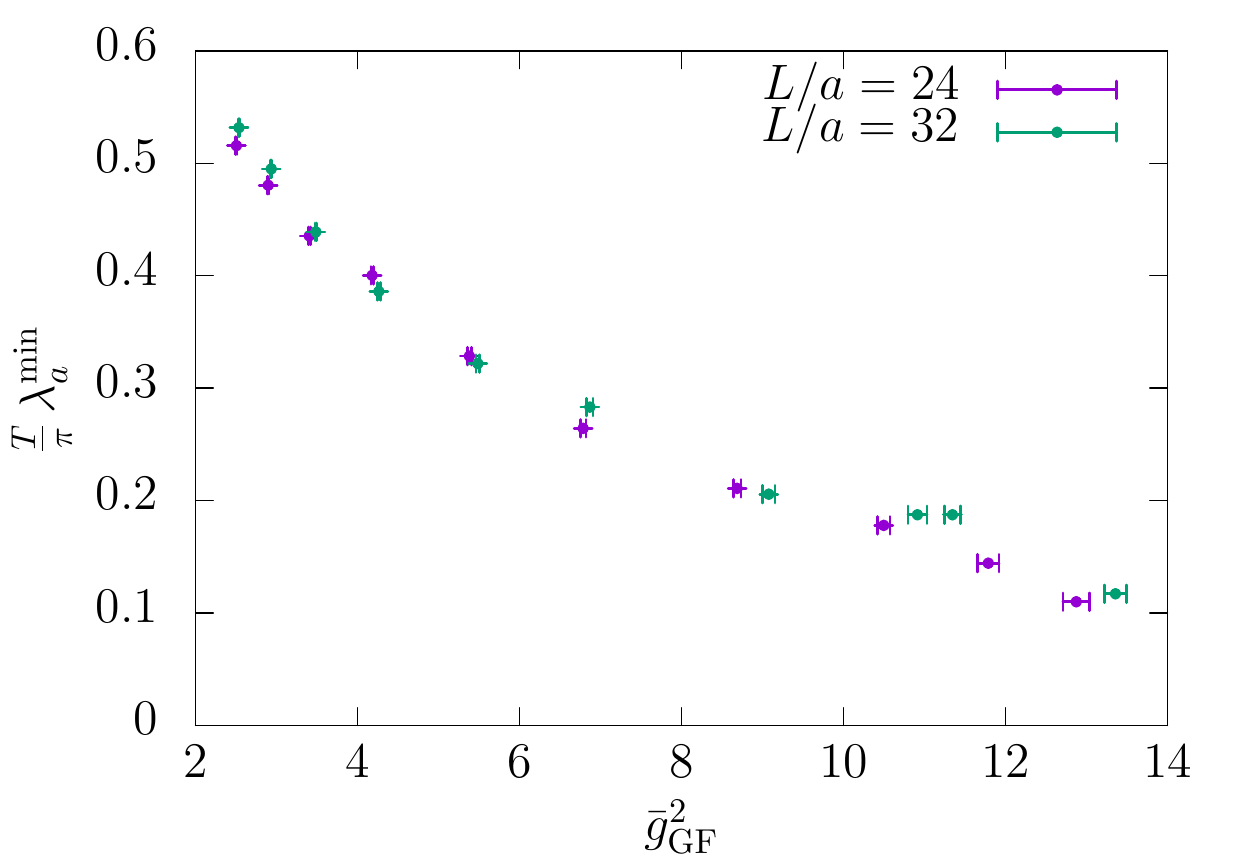}
    \caption{$T \lambda_a^{\rm min}/\pi$ as a function of $\bar g^2_{\rm GF}(L)$ for
    our larger lattices $L/a=24,32$.}
  \end{subfigure}
  \begin{subfigure}[b]{0.48\textwidth}
    \includegraphics[width=\textwidth]{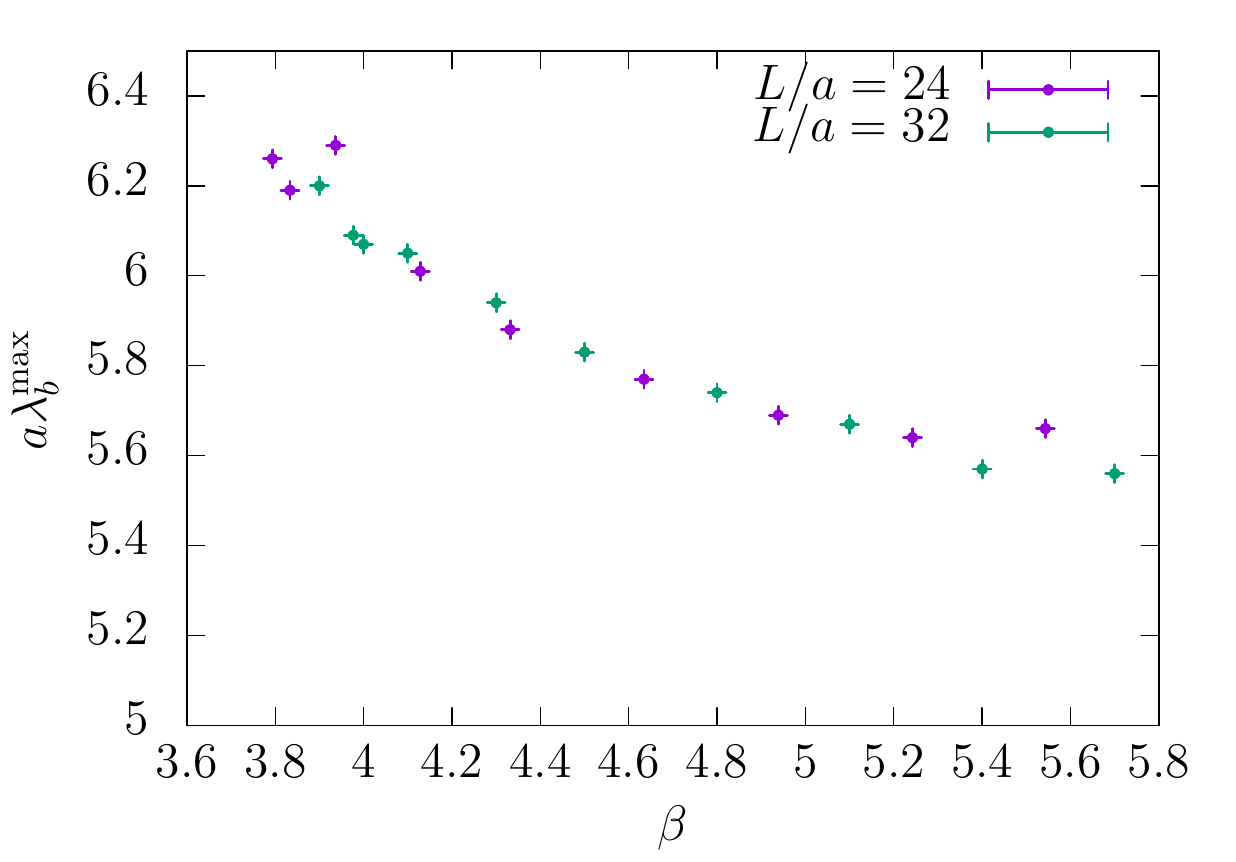}
    \caption{$a \lambda_b^{\rm max}$ as a function of $\beta$ for
    our larger lattices $L/a=24,32$.}
  \end{subfigure}

  \caption{Scaling of $\lambda_a^{\rm min}$ and $\lambda_b^{\rm max}$ determining
    the spectral range of the operator $\sqrt{\hat D^{\dagger} \hat D}$
    entering the RHMC algorithm.}
  \label{fig:rarb}
\end{figure}

\subsubsection{Rational approximation and spectral gap of the Dirac operator}
\label{a:gap}

The RHMC algorithm uses a Zolotarev approximation~\cite{zolo} in the interval $[r_a, r_b]$
for the operator $R=({\hat D}^\dag{\hat D})^{-1/2}$, which enters the decomposition,
\begin{equation}
  \label{eq:detq2}
  {\rm det}\,D =   {\rm det}(1_e + D_{oo})\,{\rm det}\,{\hat D},\qquad
  {\rm det}\,{\hat D} =   W {\rm det}\,R^{-1}.
\end{equation}
Here ${\hat D} = D_{ee}- D_{eo}D^{-1}_{oo}D_{oe}$ denotes the even-odd preconditioned
Dirac operator, and $1_e$ is the projector to the subspace of quark fields that 
vanish on the odd sites of the lattice. The operators $D_{ee}$,
$D_{eo}$, $D_{oo}$,  
and $D_{oe}$ refer to the even-even, even-odd, odd-odd, and odd-even parts
of the Dirac 
operator, respectively. The residual factor $W = {\det}(DR)$, instead, is considered as a reweighting
factor which corrects possible (small) errors in the approximation of $R$; we
estimate this using two random sources (cf. \texttt{rhmc.pdf} of the
documentation of 
the \texttt{openQCD} package for more detail information). 
The precision of our rational approximations with parameters in
\tab{tab:rarb} is very high. Consequently the reweighting taking into
account the factor $W$ has very little effect. 

Figure~\ref{fig:rarb} summarizes the values for the smallest, $\lambda_a^{\rm  min}$, 
and largest, $\lambda_b^{\rm max}$, eigenvalues of $\sqrt{{\hat D}^\dag{\hat D}}$, 
measured during our most challenging runs (those with sizes $L/a=24,32$).
More quantitative information is found in table~\ref{tab:rarb}. The main conclusion is
that even at the largest volumes, our choice of boundary conditions ensures the existence
of a gap in the Dirac operator, and with our chosen values of $r_a$, $r_b$ the simulations 
are safe. 

\begin{table}
\small
  \centering
  \begin{tabular}{cllcllcll}\toprule
    $L/a$ & $\beta$    & $\tauint$    &  $L/a$ & $\beta$    & $\tauint$       &   $L/a$ & $\beta$    & $\tauint$       \\\cmidrule(lr){1-3}\cmidrule(lr){4-6}\cmidrule(lr){7-9}
     $16$ & $3.556470$ & $60(17)^*$   &   $24$ & $3.735394$ & $175(63)^*$     &   $32$  & $3.900000$ & $   111(37)^*$  \\
     $16$ & $3.556470$ & $36(8)$      &   $24$ & $3.793389$ & $122(40)^*$     &   $32$  & $3.976400$ & $\pz 89(27)^*$  \\
     $16$ & $3.653850$ & $32(7)$      &   $24$ & $3.833254$ & $\pz 59(15)^*$  &   $32$  & $4.000000$ & $   144(50)^*$  \\
     $16$ & $3.754890$ & $26(5)$      &   $24$ & $3.936816$ & $\pz 36(8)^*$   &   $32$  & $4.100000$ & $\pz 82(22)$    \\
     $16$ & $3.947900$ & $15(2)$      &   $24$ & $4.128217$ & $\pz 36(8)$     &   $32$  & $4.300000$ & $\pz 82(17)$    \\
     $16$ & $4.151900$ & $11(2)$      &   $24$ & $4.331660$ & $\pz 30(6)$     &   $32$  & $4.500000$ & $\pz 38(7)$     \\
     $16$ & $4.457600$ & $\pz 9(1)$   &   $24$ & $4.634654$ & $\pz 17(3)$     &   $32$  & $4.800000$ & $\pz 40(7)$     \\
     $16$ & $4.764900$ & $\pz 7.8(9)$ &   $24$ & $4.938726$ & $\pz 18(3)$     &   $32$  & $5.100000$ & $\pz 34(5)$     \\
     $16$ & $5.071000$ & $\pz 7.1(8)$ &   $24$ & $5.242465$ & $\pz 14(2)$     &   $32$  & $5.400000$ & $\pz 21(3)$     \\
     $16$ & $5.371500$ & $\pz 6.2(7)$ &   $24$ & $5.543070$ & $\pz 15(8)$     &   $32$  & $5.700000$ & $\pz 23(3)$     \\
    \bottomrule
  \end{tabular}
  \caption{Integrated autocorrelation times measured in the
    simulations performed to determine the step scaling function
    $\Sigma$. 
    Measurements of $\gbar_{\rm GF}^2(L)$ with $L/a=16,24$ were separated by 
    10 MDU's, while those at $L/a=32$ are separated by 20 MDU's. 
    Accordingly some of the determined $\tau_{\rm int}$ are below one in
    units of measurements, which introduces a (small) bias. Values
    marked with an $*$ corresponds to ensembles generated with
    \texttt{Algorithm B}.}
  \label{tab:tauint}
\end{table}

\subsubsection{Scaling of autocorrelation times}

\begin{figure}
  \centering
  \includegraphics[width=0.7\textwidth]{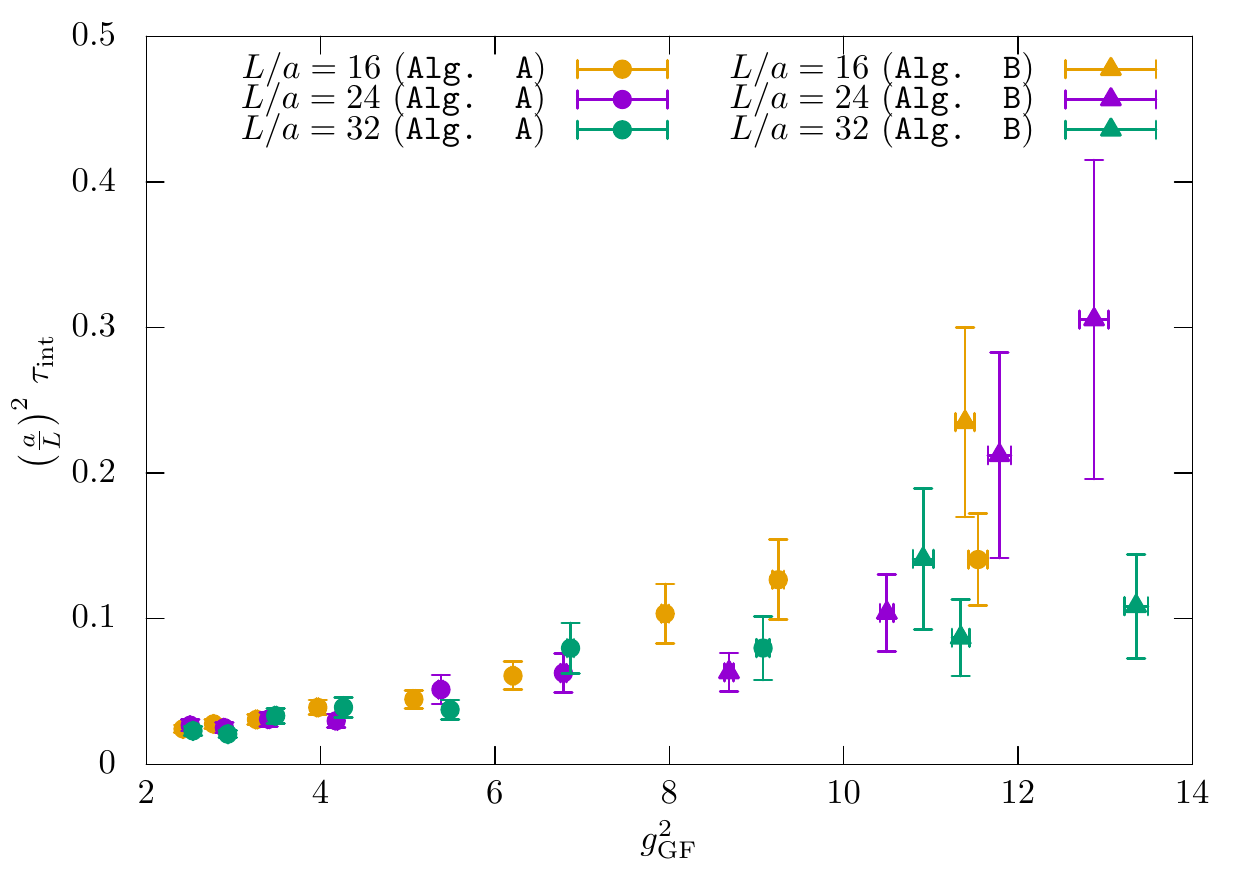}
  \caption{Scaling of $\tau_{\rm int}$ as a function of $\bar g^2(L)$
    for different $L/a$.}
  \label{fig:tauint}
\end{figure}

Once more we focus on the more challenging simulations and discuss the scaling of the 
integrated autocorrelation times in our simulations with lattice sizes $L/a=16,24,32$. 
Table~\ref{tab:tauint} shows the
autocorrelation times, determined as in \cite{Wolff:2003sm} in molecular dynamic units, while
fig.~\ref{fig:tauint} indicates that they roughly 
follow the expected scaling with $a^{-2}$ \cite{Luscher:2011kk} at
constant $\gbar_{\rm GF}^2(L)$ i.e. in fixed physical volume. 
Even the deviations from scaling seen at the larger 
coupling have a plausible explanation in terms of a 
correction to scaling. When the lattice spacing is bigger
than around 0.05~fm, the standard HMC still shows topological
activity\cite{Schaefer:2010hu}. \texttt{Algorithm B} will therefore have a  
number of attempts to change topology, which increases with the lattice spacing. These attempts are vetoed by the acceptance step,
reducing the acceptance rate and increasing the
autocorrelations. This easily explains the three 
highest lying points in the figure, but of course the
quality of the data is not good enough for a quantitative statement.

The length of our Monte Carlo chains is always between $200\,\tau_{\rm
  int}$ and $2000\,\tau_{\rm int}$.
  Despite the expectation that autocorrelations will eventually scale rather differently in large volume compared to our situation with
  Schr\"odinger functional boundary conditions, the longest autocorrelation times 
  of our finite volume simulations are comparable to the longest ones 
  observed in large volume in \cite{Bruno:2014jqa}. 

\subsubsection{The critical lines}
\label{ap:tuning}

\begin{table}[th]
  \centering
  \small
  \begin{tabular}{cccc} \toprule
    coeff.     & $L/a=8$              & $L/a=12$             & $L/a=16$             \\ \midrule
    $\zeta_0$  & $+1.005834130000000$ & $+1.002599440000000$ & $+1.001463290000000$ \\
    $\mu_0$    & $-0.000022208694999$ & $-0.000004812471537$ & $-0.000001281872601$ \\
    $\mu_1$    & $-0.202388398516844$ & $-0.201746020772477$ & $-0.201520105247962$ \\ \cmidrule(lr){1-4}
    $\zeta_1$  & $-0.560665657872021$ & $-0.802266237327923$ & $-0.892637061391273$ \\
    $\zeta_2$  & $+3.262872842957498$ & $+4.027758778155415$ & $+5.095631719496583$ \\
    $\zeta_3$  & $-5.788275397637978$ & $-6.928207214808553$ & $-8.939546687871335$ \\
    $\zeta_4$  & $+4.587959856400246$ & $+5.510985771180077$ & $+7.046607832794273$ \\
    $\zeta_5$  & $-1.653344785588201$ & $-2.076308895962694$ & $-2.625638312722623$ \\
    $\zeta_6$  & $+0.227536321065082$ & $+0.320430672213824$ & $+0.405387660384441$ \\
    $\mu_2$    & $+0.090366980657738$ & $+0.128161834555849$ & $+0.139461345465939$ \\
    $\mu_3$    & $-0.600952105402754$ & $-0.681097059845447$ & $-0.847457204378732$ \\
    $\mu_4$    & $+0.934252532135398$ & $+0.991316994385556$ & $+1.261676178806362$ \\
    $\mu_5$    & $-0.608706158693056$ & $-0.606597739050552$ & $-0.754644691612547$ \\
    $\mu_6$    & $+0.140501978953879$ & $+0.129031928169091$ & $+0.153135714480269$ \\
    \bottomrule
  \end{tabular}
  \caption{Coefficients for the parameterization
    eq.~\eqref{eq:amcresult}. The three 
    leading coefficients $\zeta_0$, $\mu_0$ in the
    upper part of the table are combinations of known perturbative
    coefficients while the  others were determined by a
    fit~\cite{Nf3tuning}. }
  \label{tab:fitparams}
\end{table}

Since we work in a massless renormalization scheme, 
we need to define and know the critical line
in the space of bare lattice parameters $(\beta,\kappa,L/a)$; or equivalently
$(g_0^2,am_0,L/a)$ with
\begin{align}\label{eq:}
   g_0^2 = 6/\beta \,, \qquad
   am_0  = (2\kappa)^{-1} - 4 \;.
\end{align}
The critical line, $am_0=a\mcr(g_0,a/L)$, is defined by 
$m_1=0$, where  $m_1$ is a current quark mass in an 
$(L/a)^4$ lattice. Making $\mcr$ dependent on $L/a$ in this way
and using the same $L/a$ in this definition as in $\Sigma$, 
the cutoff effects are guaranteed to disappear as $\rmO(a^2)$ in the 
improved theory. Details on $m_1$ as well as on the many
precise simulations done to find the critical lines by interpolation
can be found in~\cite{Nf3tuning}. 

For completeness we here list 
the results needed to compute $\mcr$. 
In \Tab{tab:fitparams} we provide the coefficients of
the interpolating functions for the critical lines, 
\begin{align}\label{eq:amcresult}
        a\mcr(g_0,a/L) &= \left( {\sum}_{k=0}^{6} \mu_k\, g_0^{2k} \right) \times \left( {\sum}_{i=0}^{6} \zeta_i \, g_0^{2i} \right)^{-1} \,,
\end{align}
at a given value of $L/a$, valid for all values of $g_0^2$ used in
this paper. 
These parameterizations guarantee $m_1 L < 0.005$. With
these coefficients the reader can reconstruct  
the input mass-parameter $\kappa_{\rm cr}$ corresponding to our 
simulations. 

\vskip1em

\subsection{Tuning to selected couplings.}
\label{ap:tuning_gbar}
In \tab{tab:smallL} we collect our raw data for $\gbar^2_{\rm GF}(L)$ 
on the small lattices.

\input{tables/raw_tuning.tex}

As explained in the main text, we make maximum use of these data by
performing smooth interpolations for $L/a=8,12$. This enables a very
precise determination of $\gbar_{\rm GF}^2(L)$ for those bare parameters where
we have computed $\gbar_{\rm GF}^2(2L)$. 

At fixed $L/a$ we fit 
\begin{equation}
  v(\beta) = 1/\gbar^2_{\rm GF}(L)\,,
\end{equation}
to a Pad\'e  ansatz of degrees $[n_1,n_2]$,
\begin{equation}
  v(\beta) = \frac{\sum_{n=0}^{n_1} a_n\beta^n}{1 + \sum_{n=1}^{n_2} b_n\beta^n}\,,
\end{equation}
and obtain predictions $\gbar_{\rm GF}^2(L)$ at the desired $\beta$ from the fit
and their errors from the covariance matrix of the fit parameters.

\begin{figure}
  \centering
  \includegraphics[width=0.8\textwidth]{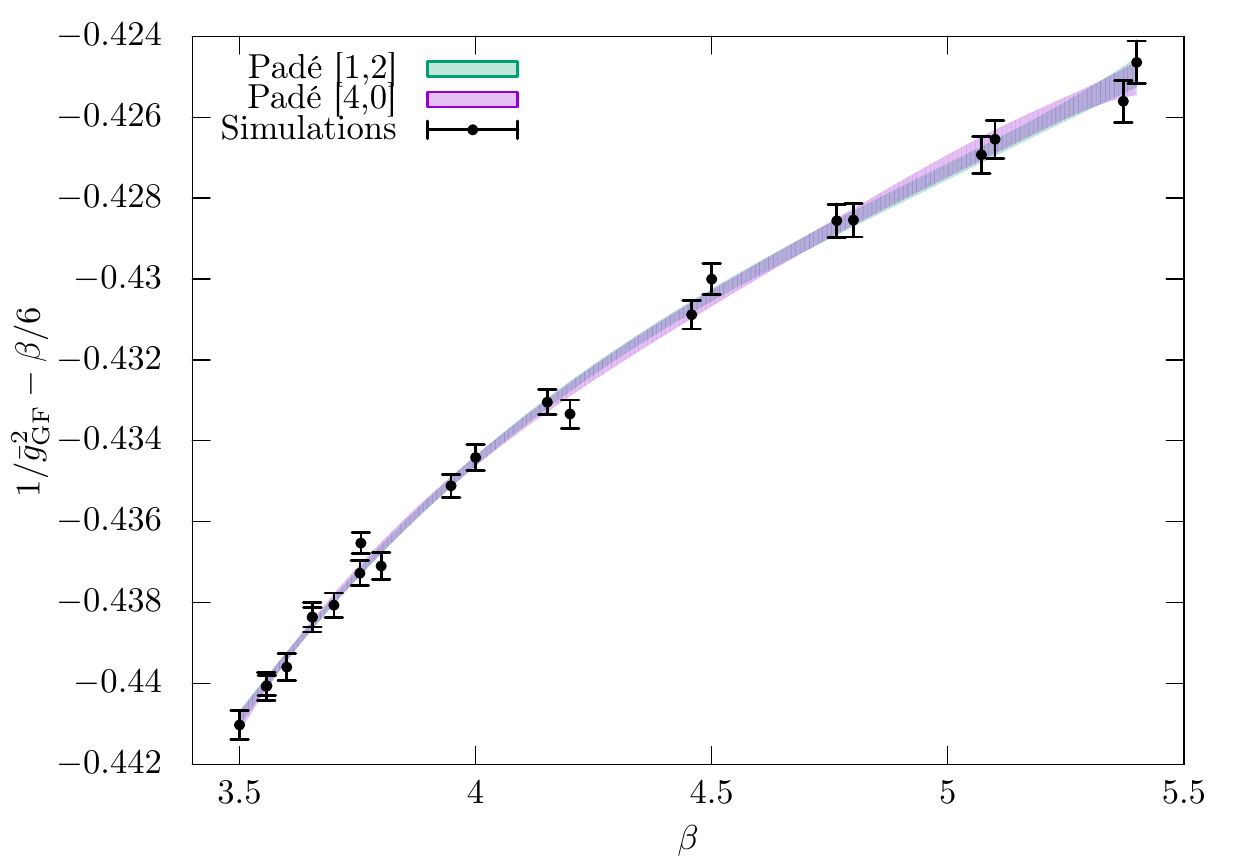}
  \caption{$\frac{1}{\gbar_{\rm GF}^2}-\frac{\beta}{6}$ as a function of
    $\beta$ for $L/a=8$. The simulation points 
    are fitted to a $[1,2]$ Pad\'e ($\chi^2=17.05$ for 18
    degrees of freedom) and a  $[4,0]$ Pad\'e  
    ($\chi^2=15.42$ for 17 degrees of freedom). 
    The different fit functions are hard to distinguish.}
  \label{fig:fit8}
\end{figure}

In fig.~\ref{fig:fit8} we show a couple of typical fits of all the
$L/a=8$ data to a $[4,0]$ Pad\'e and a  $[1,2]$ one. These
fits have a good quality. Other 
fit functions were tested 
with the result that, once the fits have a reasonable 
number of degrees of freedom and a good $\chi^2$, 
the interpolated values of $\gbar_{\rm GF}^2(L)$ 
are entirely stable within their errors.
This holds also for the $L/a=12$ data. As final description of our
 $L/a=8,12$ data we use the 
$[4,0]$ and $[3,0]$ Pad\'e, i.e. a simple polynomial of degree three and four, respectively. These choices
 yield the
 values of $\gbar_{\rm GF}^2(L)$ listed in \tab{tab:Sigma}.

%% file: tables/raw_tuning.tex
\begin{table}[th]
  \small
  \centering
  \begin{tabular}{ccccrcccccrl}  \toprule
    $L/a$ & $\beta$   & $\gbar^2$            & $\Nms$ & $N_Q$& $L/a$& $\beta$   & $\gbar^2$        & $\Nms$ & $N_Q$  \\ \cmidrule(lr){1-5}\cmidrule(lr){6-10}
     $8 $ & $3.50000$ & $\pz 7.0271(180)$    & $5001$ & $55$ & $8$  & $4.00000$ & $4.3057(\pz 60)$ & $5001$ &  $0$   \\ 
     $8 $ & $3.55647$ & $\pz 6.5501(149)$    & $5001$ & $34$ & $8$  & $4.15190$ & $3.8619(\pz 45)$ & $5001$ &  $0$   \\ 
     $8 $ & $3.55800$ & $\pz 6.5385(106)$    & $5001$ & $20$ & $8$  & $4.20000$ & $3.7501(\pz 49)$ & $5001$ &  $0$   \\ 
     $8 $ & $3.60000$ & $\pz 6.2343(131)$    & $5001$ & $35$ & $8$  & $4.45760$ & $3.2046(\pz 36)$ & $5001$ &  $0$   \\ 
     $8 $ & $3.65385$ & $\pz 5.8612(126)$    & $5001$ & $ 7$ & $8$  & $4.50000$ & $3.1250(\pz 37)$ & $5001$ &  $0$   \\ 
     $8 $ & $3.65452$ & $\pz 5.8574(\pz 84)$ & $5001$ & $ 7$ & $8$  & $4.76490$ & $2.7353(\pz 31)$ & $5001$ &  $0$   \\ 
     $8 $ & $3.70000$ & $\pz 5.5990(\pz 93)$ & $5001$ & $ 1$ & $8$  & $4.80000$ & $2.6921(\pz 30)$ & $5001$ &  $0$   \\ 
     $8 $ & $3.75489$ & $\pz 5.3040(\pz 88)$ & $5001$ & $ 0$ & $8$  & $5.07100$ & $2.3910(\pz 26)$ & $5001$ &  $0$   \\ 
     $8 $ & $3.75709$ & $\pz 5.2728(\pz 74)$ & $5001$ & $ 0$ & $8$  & $5.10000$ & $2.3615(\pz 26)$ & $5001$ &  $0$   \\ 
     $8 $ & $3.80000$ & $\pz 5.0959(\pz 87)$ & $5001$ & $ 0$ & $8$  & $5.37150$ & $2.1293(\pz 24)$ & $5001$ &  $0$   \\ 
     $8 $ & $3.94790$ & $\pz 4.4870(\pz 56)$ & $5001$ & $ 0$ & $8$  & $5.40000$ & $2.1037(\pz 23)$ & $5001$ &  $0$   \\\cmidrule(lr){1-5}\cmidrule(lr){6-10}
     $12$ & $3.40000$ & $11.3081(994)$       & $5000$ &$\es$ & $12$ & $4.33166$ & $3.8725(\pz 60)$ & $5001$ &  $0$   \\ 
     $12$ & $3.50000$ & $\pz 9.1035(284)$    & $5000$ &$\es$ & $12$ & $4.50000$ & $3.4738(\pz 54)$ & $5001$ &  $0$   \\ 
     $12$ & $3.70000$ & $\pz 6.8400(167)$    & $5001$ & $69$ & $12$ & $4.63465$ & $3.2051(\pz 47)$ & $5001$ &  $0$   \\ 
     $12$ & $3.73539$ & $\pz 6.5428(176)$    & $5001$ & $18$ & $12$ & $4.80000$ & $2.9255(\pz 32)$ & $8000$ &  $0$   \\ 
     $12$ & $3.80000$ & $\pz 6.0832(123)$    & $5001$ & $ 8$ & $12$ & $4.93873$ & $2.7371(\pz 38)$ & $5001$ &  $0$   \\ 
     $12$ & $3.83325$ & $\pz 5.8685(134)$    & $5001$ & $ 3$ & $12$ & $5.10000$ & $2.5470(\pz 26)$ & $8000$ &  $0$   \\ 
     $12$ & $3.90000$ & $\pz 5.4794(106)$    & $5001$ & $ 2$ & $12$ & $5.24247$ & $2.3919(\pz 25)$ & $8000$ &  $0$   \\ 
     $12$ & $3.93682$ & $\pz 5.2996(107)$    & $5001$ & $ 1$ & $12$ & $5.40000$ & $2.2394(\pz 22)$ & $8000$ &  $0$   \\ 
     $12$ & $4.00000$ & $\pz 4.9991(100)$    & $5001$ & $ 0$ & $12$ & $5.54307$ & $2.1213(\pz 21)$ & $8000$ &  $0$   \\ 
     $12$ & $4.12822$ & $\pz 4.4945(\pz 75)$ & $5001$ & $ 0$ & $12$ & $5.60000$ & $2.0823(\pz 21)$ & $8001$ &  $0$   \\ 
     $12$ & $4.20000$ & $\pz 4.2480(\pz 66)$ & $5001$ & $ 0$ &                                                       \\\cmidrule(lr){1-5}\cmidrule(lr){6-10}
     $16$ & $3.90000$ & $\pz 6.5489(155)$    & $4600$ & $15$ & $16$ & $4.80000$ & $3.2029(\pz 52)$ & $5000$ &  $0$   \\ 
     $16$ & $4.00000$ & $\pz 5.8673(140)$    & $4602$ & $35$ & $16$ & $5.10000$ & $2.7359(\pz 35)$ & $6001$ &  $0$   \\ 
     $16$ & $4.10000$ & $\pz 5.3013(134)$    & $3200$ &  $0$ & $16$ & $5.40000$ & $2.3900(\pz 30)$ & $6001$ &  $0$   \\ 
     $16$ & $4.30000$ & $\pz 4.4901(\pz 77)$ & $5000$ &  $0$ & $16$ & $5.70000$ & $2.1257(\pz 25)$ & $7001$ &  $0$   \\ 
     $16$ & $4.50000$ & $\pz 3.8643(\pz 63)$ & $5000$ &  $0$ & $16$ & $3.97640$ & $6.0369(142)$ & 4567 & 0     \\
    \bottomrule
  \end{tabular}
  \caption{Coupling results on the small lattices for various $\beta=6/g_0^2$
           and $L/a$.  The separation of measurements is $5-10$ MDU.  $\Nms$
           denotes the number of measurements out of which $N_Q$ have non-zero
           charge $Q$. The effective number of measurements is $\Nms - N_Q$.
           Simulations with $N_Q=\emptyset$ were carried out with 
           \texttt{Algorithm B}. 
          }
  \label{tab:smallL}
\end{table}
